\documentclass[amsmath,amssymb,aps,prd,twocolumn,superscriptaddress,10pt]{revtex4-1}
\usepackage{graphicx}
\usepackage{dcolumn}
\usepackage{subfigure}
\usepackage{color}
\usepackage{footnote}
\usepackage{url}

\newcommand\apjl{The Astrophysical Journal Letters}
\newcommand\aj{The Astronomical Journal}
\newcommand\mnras{Monthly Notices of the Royal Astronomical Society}
\newcommand\aap{Astronomy and Astrophysics}
\newcommand\jcap{Journal of Cosmology and Astroparticle Physics}
\newcommand\araa{Annual Review of Astronomy and Astrophysics}

\begin{document}

\title{Preliminary study on parameter estimation accuracy of supermassive black hole binary inspirals
for TianQin}

\author{Wen-Fan Feng}
\affiliation{MOE Key Laboratory of Fundamental Physical Quantities Measurements,
Hubei Key Laboratory of Gravitation and Quantum Physics, PGMF and
School of Physics, Huazhong University of Science and Technology,
Wuhan 430074, China}

\author{Hai-Tian Wang}
\affiliation{TianQin Research Center for Gravitational Physics, Sun Yat-Sen University,
Zhuhai 519082, China}
\affiliation{School of Physics and Astronomy, Sun Yat-Sen University,
Zhuhai 519082, China}

\author{Xin-Chun Hu}
\affiliation{MOE Key Laboratory of Fundamental Physical Quantities Measurements,
Hubei Key Laboratory of Gravitation and Quantum Physics, PGMF and
School of Physics, Huazhong University of Science and Technology,
Wuhan 430074, China}

\author{Yi-Ming Hu}
\email{huyiming@sysu.edu.cn }
\affiliation{TianQin Research Center for Gravitational Physics, Sun Yat-Sen University,
Zhuhai 519082, China}
\affiliation{School of Physics and Astronomy, Sun Yat-Sen University,
Zhuhai 519082, China}

\author{Yan Wang}
\email{ywang12@hust.edu.cn}
\affiliation{MOE Key Laboratory of Fundamental Physical Quantities Measurements,
Hubei Key Laboratory of Gravitation and Quantum Physics, PGMF and
School of Physics, Huazhong University of Science and Technology,
Wuhan 430074, China}

\date{\today}

\begin{abstract}

  We use the Fisher information matrix method to calculate the parameter estimation
  accuracy of inspiraling supermassive black holes binaries for TianQin, a proposed 
  space-borne laser interferometric detector aimed at detecting gravitational
  waves in the millihertz frequency band.
  The `restricted' post-Newtonian waveform in which third order
  post-Newtonian (3PN) phase including spin effects (spin-orbit $\beta$ and spin-spin $\sigma$)
  and first-order eccentricity contribution is employed.
  Monte Carlo simulations using $10^3$ binaries for mass pairs with
  component masses in the range of $({10^5},{10^7}){M_ \odot }$
  and cosmological redshift $z=0.5$ show that the medians of the root-mean-square error distributions
  for the chirp mass $M_c$ and
  symmetric mass ratio $\eta$ are in the range of $\sim 0.02\% - 0.7\% $ and
  $\sim 4\% - 8\% $, respectively. The luminosity
  distance $D_L$ can be determined to be $\sim 1\% - 3\% $, and
  the angular resolution of source $\Delta \Omega $ is better than 12 ${\deg ^2}$.
  The corresponding results for $z=1.0$ and $2.0$, which are deteriorated with the
  decreasing of the signal-to-noise ratio, have also been given.
  We show that adding spin parameters degrades measurement accuracy of the
  mass parameters (${M_c}$, $\eta$), and the time and the orbital
  phase of coalescence ($t_c$, $\phi _c$); the inclusion of the first-order eccentricity
  correction to the phase worsens the estimation accuracy comparing with the circular cases.
  We also show the effects of post-Newtonian order on parameter estimation accuracy by comparing
  the results based on second order and third order post-Newtonian phases.
  Moreover, we calculate the horizon distance
  of supermassive black hole binaries for TianQin.

\end{abstract}

\pacs{}
\maketitle

\section{Introduction}

The observational window of gravitational wave (GW) astronomy has been opened by 
the landmark detections by the Laser Interferometer Gravitational-Wave 
Observatory (LIGO) and Virgo \cite{2016PhRvL.116f1102A, 2016PhRvL.116x1103A, 2017PhRvL.119n1101A, 
2017PhRvL.118v1101A, 2017ApJ...851L..35A, 2017PhRvL.119p1101A, 2018arXiv181112907T}. 
Due to the limitations from gravity gradient noise and seismic noise, 
it is extremely challenging to detect GWs with frequencies $< 10$ Hz by 
the terrestrial interferometers. 
In the low frequency band, the Laser Inteferometer Space Antenna (LISA) \cite{2017arXiv170200786A}
had been selected as the third large-class mission of the European Space Agency (ESA) 
with an anticipated launch around 2034. 
The key technologies for LISA, such as gravity reference system and space laser 
interferometry, have been successfully demonstrated by the LISA Pathfinder 
\cite{PhysRevLett.116.231101}.

TianQin is a proposed space-borne laser interferometric detector for gravitational
waves in millihertz frequencies (0.1 mHz -1 Hz) \cite{Luo2016}.
The detector comprises three
identical drag-free satellites orbiting around the Earth in a nearly equilateral
triangular constellation.
The geocentric distance of each satellite is approximately ${10^5}$ km which
makes the arm length of the interferometer be about $1.73 \times {10^5}$ km.
The Keplerian orbit's period of each satellite due to the gravitational attraction 
from the Earth is approximately $3.65$ days. 
Each pair of satellites is interconnected by two-way infrared laser beams
which forms up to three Michelson interferometers.
The guiding center of the constellation coincides with the geocenter
and it moves around the Sun in the ecliptic orbit.
The normal of the detector
plane points toward the tentative reference source RX J0806+15, which is among
the strongest GW sources in Galactic ultra-compact white-dwarf binaries
\cite{Israel2002,Strohmayer2008,2019tobesubHuang}. The fundamentals of the satellite orbits and the response
of each Michelson interferometer for TianQin have been studied in \cite{Hu2018}.
The nominal orbits and a set of alternative orbits have been
optimized such that the stability requirements on arm length variation, relative
velocity and breathing angle of the triangular constellation can be satisfied
for a five-year mission lifetime \cite{doi:10.1142/S0218271819501219}.

As a millihertz frequency gravitational wave observatory, the design of the architecture for
the TianQin mission and the trade-off among a variety of technologies for the instruments
are driven by the attainable science objectives.
The main categories of GW sources in TianQin's frequency band \cite{Hu2017} are
Galactic ultra-compact binaries, coalescing supermassive black hole binaries
(SMBHBs), capture of stellar-mass compact objects by
a massive black hole (MBH), i.e., extreme-mass-ratio inspirals (EMRIs) \cite{Hughes2001}, 
inspiral of stellar mass black hole binaries \cite{PhysRevLett.116.231102},
and stochastic GW background originating from primordial BHs \cite{2018JCAP...07..007D}
and/or cosmic strings \cite{2010PhRvD..81j4028O}.
Among these, SMBHB mergers are arguably the most powerful
GW sources; therefore, they deserve detailed investigations.

Observations indicate that the center of almost every galaxy
hosts a SMBH whose mass is in the range of $10^{5}-10^{10}{M_ \odot}$
\cite{1995ARA&A..33..581K}.
In the hierarchical merger scenario of galaxy formation, large galaxies are assembled through
multiple mergers during their lifetime. As they merge, the center SMBHs will
approach each other due to the deep gravitational potential, dynamic friction, triple interaction,
gravitational waves, etc. \citep{2003AIPC..686..201M,2003AIPC..686..161K}. 
SMBHB can even be formed directly inside the first galaxies
under certain physical conditions \citep{2003ApJ...596...34B}.
SMBHB merger rates along the cosmic
history predicted by three different population models and the science capability
for different eLISA designs have been discussed \cite{Klein2016}.
The work on estimating SMBHB detection rates for TianQin based on the
semi-analytical model is underway \cite{2019arXiv190204423W}. 
The science potential of TianQin on testing the black hole 
no-hair theorem by using the ringdown signals from SMBHB mergers 
has been studied \cite{2019arXiv190208922S}.

In addition to detection rates, it is important to forecast how
accurately that TianQin can measure the parameters pertinent to
SMBHB systems which potentially can be used to enable the subsequent
studies, for example, formation and growth mechanism of
seed black holes \cite{2001ApJ...551L..27M,2003ApJ...596...34B},
co-evolution of SMBHBs with host galaxies \cite{1998AJ....115.2285M,2013ARA&A..51..511K}
and cosmography \cite{2003CQGra..20S..65H,2016JCAP...04..002T}.
From the perspective of data analysis, one needs \textit{matched filtering}
to extract the deterministic signals, such as the ones from inspiral stage of
SMBHBs, from the noisy data. This involves passing the detector's
strain data output through a bank of templates that are characterized by
the signal parameter set \citep{1998.book.....KayII,1968.book.....H}.
In general, Monte-Carlo simulation using synthetic data and a data-analysis pipeline
will be required to systematically evaluate the performance of parameter estimation
and signal reconstruction for a specific detector.
However, this procedure is cumbersome and computationally demanding.
For the ease of use and computational efficiency, 
the Fisher information matrix (FIM) method has been implemented to bound the
parameter estimation accuracy of post-Newtonian (PN) inspirals for both
stellar-mass black hole binaries detected by the ground-based detector LIGO \cite{Cutler1994}
and SMBHBs detectable by the space-borne detector LISA \cite{Cutler1998}.
It has been shown that the results based on the FIM method are consistent
with the ones from more sophisticated Bayesian parameter
estimations for high signal-to-noise ratio (SNR) cases \cite{Rodriguez2013,Porter2015}.

In the applications of FIM method, Cutler and Flanagan \cite{Cutler1994} used
the \textit{restricted} 1.5PN inspiral waveform that includes spin-orbit parameter ($\beta$) 
to discuss the estimation accuracy of luminosity distance $D_L$, chirp mass $M_c$,
reduced mass $\mu$ and $\beta$ for LIGO and Virgo network.
By extending the waveform template phasing to 2PN and
including both spin-orbit ($\beta$) and spin-spin ($\sigma$) parameters,
Poisson and Will \cite{Poisson1995} found that the 1.5PN phasing actually underestimated
the root-mean-square (rms) errors in $M_c$, $\mu$, and $\beta$.
Furthermore, Arun et al. \cite{Arun2005,Arun2006} adopted 3.5PN phasing,
however, ignored spin and eccentricity effects, to emphasize the importance
of employing higher order PN correction for parameter estimation.

For LISA, Cutler \cite{Cutler1998} first used the waveform in \cite{Cutler1994}
to calculate the sky location error, which was later extended to 2PN waveform
including spin effects by Berti et al. \cite{Berti2005}.
Lang et al. \cite{Lang2006,Lang2011} used 2PN waveform including the precession-induced
modulations with partially aligned spins and higher PN harmonics to
examine the impact upon parameter estimation for LISA. They found that the
additional precession periodicity and higher PN harmonics can improve 
the accuracy. 
The first-order phase correction due to orbital eccentricity has been considered
in \cite{Krolak1995}. Afterwards, it has been found that high eccentricity ($e_0 \geq 0.6$)
not only enhances the SNR but also improves LISA's angular resolution of SMBHBs
than the circular obits for relatively high equal mass systems
($\sim 10^{7} M_{\odot} + 10^{7} M_{\odot}$) \cite{Mikoczi2012,Gondan2017}.

In the current work, we mainly focus on the investigation
of parameter estimation accuracy for SMBHB inspirals that can be achieved by TianQin.
We employ the restricted post-Newtonian waveform that includes up to
third order (3PN) phase and contributions from spin effects (both spin-orbit and spin-spin) and
first-order eccentricity effects. 
The plan for the rest of the paper is as follows. In Sec.~\ref{sec:waveform_model},
we give the GW waveform used in this work.
In Sec.~\ref{sec:detection}, we briefly summarize the characteristics of the TianQin detector
and its prospects for detecting GWs from inspiraling SMBHBs.
In Sec.~\ref{sec:FIM}, we review some basics of the FIM method.
Sec.~\ref{sec:simulation} carries out the Monte-Carlo simulations
for typical SMBHBs and presents the main results of the rms error distributions of
estimated parameters and the comparison with variations in employed waveforms.
Conclusions and discussions on possible future studies
are given in Sec.~\ref{sec:conclusion}.
In Appendix~\ref{sec:SPA}, we provide the details of 
the transforming GW signal in time-domain to
frequency-domain based on the stationary phase approximation (SPA),
which leads to validating the requirements of SPA for TianQin in Appendix~\ref{sec:test}.
Throughout this paper we adopt the geometrical units in which $G = c = 1$.

\section{Waveform model}\label{sec:waveform_model}

\subsection{The time-domain waveform}\label{subsec:twaveform}

In the source rest frame, we construct the center-of-mass coordinates
$\lbrace x_1,x_2,x_3 \rbrace$ and consider a binary
consisting of two masses $m_1$ and $m_2$ in a circular orbit
on the $x_1-x_2$ plane.
We assume that there is no precession; thus, the orbital angular momentum
vector ${\bf L}$ points along a fixed direction, i.e. the $x_3$ axis
(here we adopt the conventions in \cite{Apostolatos1994,Creighton2011}).
The source locates at a distance of $r$ with an orbital inclination angle $\iota$ defined
as $\cos \iota  = - \bf \hat L \cdot \bf \hat N$, where ${\bf \hat N}$ is the
unit vector pointing towards the source from the detector and ${\bf \hat L}$
is the unit vector of ${\bf L}$. Using the Newtonian mass quadrupole formula,
we can obtain the time-domain waveforms for the two polarizations of the GWs
propagating along the ${\bf -\hat N}$ direction:
\begin{widetext}
\begin{subequations}
\begin{eqnarray}
  {h_ + }(t) &=&  - \frac{{{M_c}}}{r}\frac{{1 + {{\cos }^2}\iota }}{2}{\left( {\frac{{{t_c} - t}}{{5{M_c}}}} \right)}^{-1/4}
   \cos \left[ {2{\varphi _c} - 2{\left( {\frac{t_c - t}{5{M_c}}} \right)}}^{5/8} \right], \\
  {h_ \times }(t) &=&  - \frac{{{M_c}}}{r}\cos \iota {\left( {\frac{{{t_c} - t}}{{5{M_c}}}} \right)}^{-1/4}
  \sin \left[ {2{\varphi _c} - 2{\left( {\frac{t_c - t}{5{M_c}}} \right)}}^{5/8} \right],
\end{eqnarray}\label{eq:waveform}
\end{subequations}
\end{widetext}
where $M_{c} = \mu^{3/5}M^{2/5}=\eta^{3/5}M$ is the chirp mass (the total mass
$M = {m_1} + {m_2}$, the reduced mass $\mu = {m_1}{m_2}/M$ and the symmetric mass
ratio ${\eta  = \mu /M }$), $t_c$ and ${\varphi _c}$ are the time and the
orbital phase of coalescence, respectively.

For detecting continuous GW signals by a space-borne detector,
it may be more convenient to work in the heliocentric ecliptic frame.
The GW strain $h(t)$ depends on the detector's antenna response
to the two GW polarizations ${h_{+,\times}}(t)$,
\begin{equation}
 h(t) = {F_ + }(\theta ,\phi ,\psi ){h_ + }(t) + {F_ \times }(\theta ,\phi ,\psi ){h_ \times }(t)  \,.
\end{equation}
Here $F_{+,\times}$ are the \textit{antenna pattern functions}.
$(\theta,\phi )$ denotes the source's ecliptic colatitude and longitude,
$\psi $ is the polarization angle between one of the semi-major axes
of the ellipse of GW polarization and the line of nodes.
The orbit and orientation of the detector determine the antenna response to
the incoming GWs. For the preliminary concept of TianQin \cite{Luo2016},
the response of a Michelson-type interferometer
that is valid in the full range of the interested frequencies
has been given in \cite{Hu2018}. Specifically, the antenna pattern
functions can be written as follows: 
\begin{subequations}
\begin{eqnarray}
 F_{+}(t) &=& {D_ + }(t,f)\cos (2\psi ) - {D_ \times }(t,f)\sin (2\psi )  \,, \\
 F_{\times }(t) &=& {D_ + }(t,f)\sin (2\psi ) + {D_ \times }(t,f)\cos (2\psi )  \,.
\end{eqnarray}
\end{subequations}
In the low-frequency region ($f<f_{\ast}\approx 0.28$ Hz) that is most relevant
to the GWs from an inspiraling SMBHB, $F_{+,\times}$ becomes independent
of the GW frequency $f$, and $D_{+,\times}$ can be analytically approximated
as \cite{Hu2018}
\begin{widetext}
\begin{subequations}
\begin{eqnarray}
  {D_+}(t;\theta,\phi) =&& \frac{{\sqrt 3 }}{{32}}\bigg(4\cos (2{\kappa _1})\Big(\big(3 + \cos(2\theta)\big)\cos{\bar \theta}\sin(2\phi - 2{\bar \phi})
    + 2\sin(\phi - {\bar \phi})\sin(2\theta)\sin({\bar \theta}) \Big)  \nonumber\\
  && -\sin(2{\kappa _1})\Big(3 + \cos(2\phi - 2{\bar \phi})\big(9 + \cos(2\theta)(3 + \cos(2{\bar \theta}))\big)
  - 6\cos(2{\bar \theta})\sin^2(\phi - {\bar \phi})  \nonumber\\
  &&- 6\cos(2\theta)\sin^2({\bar \theta})
   + 4\cos(\phi - {\bar \phi})\sin(2\theta)\sin(2{\bar \theta})\Big)\bigg)  \,,  \\
  {D_\times}(t;\theta,\phi) =&& \frac{{\sqrt 3 }}{8}\bigg(-4\cos(2{\kappa _1})\Big(\cos(2\phi - 2{\bar \phi})\cos(\theta)\cos({\bar \theta})
   + \cos(\phi - {\bar \phi})\sin(\theta)\sin({\bar \theta})\Big) \nonumber\\
   &&+ \sin(2{\kappa _1}) \Big(-\cos(\theta)\big(3 + \cos(2{\bar \theta})\big)\sin(2\phi - 2{\bar \phi})
   - 2\sin(\phi - {\bar \phi})\sin(\theta)\sin(2{\bar \theta})\Big)\bigg)  \,.
\end{eqnarray}
\end{subequations}
\end{widetext}
Here $\kappa _{1} = 2\pi f_{\rm{sc}}t + \kappa _0$, $f_{\rm{sc}}=1/(3.65$ day),
and $\kappa _0$ is the constant phase determined by the setup of
the satellites' coordinates (see \cite{Hu2018} for details).
$(\bar \theta = 1.65, \bar \phi = 2.10)$ are the colatitude and
longitude of the reference source RX J0806+15 in the
heliocentric-ecliptic frame towards which the normal of the
detector's plane is pointed \cite{Luo2016}.

As we will see in Sec.~\ref{sec:detection}, TianQin can detect SMBHBs
located at cosmological distances; therefore, it is natural to replace the distance
variable $r$ in Eq.~(\ref{eq:waveform}) by luminosity distance $D_L$.
In the standard flat $\Lambda$CDM 
cosmological model, $D_L$ can be expressed as a function
of the cosmological redshift $z$ of the source as
\begin{equation}
 D_{L} = \frac{1 + z}{H_{0}}\int_0^z \frac{dz'}{\sqrt {\Omega _{M}
 (1 + z')^3 + \Omega _{\Lambda }} }    \,,
 \label{DL_z}
\end{equation}
where the matter density $\Omega_M = 0.32$, the dark energy
density $\Omega_\Lambda = 0.68$, and Hubble constant
${H_0} = 67~{\rm km~s}^{-1}{\rm Mpc}^{-1}$ \cite{Calabrese2017}.
Correspondingly, the chirp mass and total mass measured in the
source rest frame can be related to their redshifted counterparts by
\begin{subequations}
\begin{eqnarray}
 {M'_c} &=& (1 + z){M_c}  \,, \\
 M' &=& (1 + z)M \,.
\end{eqnarray}\label{redmass}
\end{subequations}
For simplicity, hereafter we will omit the prime symbol and redefine ${M_c}$
and $M$ as the \textit{redshifted} chirp mass and total mass that are
measured in the detector's frame unless otherwise specified.

\subsection{The frequency-domain waveform}

In this subsection we compute the Fourier transform of the time-domain GW signal.
We extend Eq.~(\ref{eq:waveform}) to the ``restricted" PN waveform for which
the amplitude is kept to the leading order quadrupole term while the phase
is kept to higher PN orders, since it is more important to make the phase
coherent in GW signal detection.

Given a GW strain signal ${h(t)=A(t){\cos \Phi (t)}}$, we can obtain its
Fourier transform $\tilde h(f)$ analytically using the stationary phase
approximation (SPA, see Appendix~\ref{sec:SPA} for details)
under required constraints (see Appendix~\ref{sec:test} for the validation)
as follows,
\begin{eqnarray}\label{eq:hf}
  \tilde h(f) = AQf^{-7/6}{e^{\mathrm{i} \Psi (f)}} \,,  \,  {\rm{for}} ~ f>0  \,,
\end{eqnarray}
with $\mathrm{i}^2 = -1$ and
\begin{equation} \label{eq:amp}
 A =  - \sqrt {\frac{5}{96}} \frac {{M_c}^{5 /6}} {\pi ^{2/3}{D_L}}  \,,
\end{equation}
and
\begin{equation}\label{eq:Qfac}
  Q = \sqrt{ (1 + \cos^2{\iota})^2 F^2_{+}\big(t(f)\big) + (2\cos\iota)^2 F^2_{\times} \big(t(f)\big) }  \,.
\end{equation}
Here and hereafter, we express $t$ as a function of $f$
in the frequency-domain waveform. For the phase of waveform
that includes 3PN, spin and eccentricity effects,
$t(f)$ can be explicitly written as \cite{Buonanno2009,Krolak1995}
\begin{equation}\label{eq:time}
   t(f) = {t_c} - \frac{5}{{256}}{{ M_c }^{{{ - 5} \mathord{\left/
   {\vphantom {{ - 5} 3}} \right.
   \kern-\nulldelimiterspace} 3}}}{\left( {\pi f} \right)^{{{ - 8} \mathord{\left/
   {\vphantom {{ - 8} 3}} \right.
   \kern-\nulldelimiterspace} 3}}}\sum\limits_{k = 0}^6 {{\tau _k }{x^{{k  \mathord{\left/
   {\vphantom {\alpha  2}} \right.
   \kern-\nulldelimiterspace} 2}}}}  + {\tau_e}    \,,
\end{equation}
with the coefficients
\begin{subequations}
\begin{eqnarray}
  {\tau_0} =&& 1   \,,  \\
  {\tau_1} =&& 0  \,, \\
  {\tau_2} =&& \frac{4}{3}\left( {\frac{{743}}{{336}} + \frac{{11}}{4}\eta } \right)  \,,  \\
  {\tau_3} =&&  - \frac{8}{5}\left( {4\pi  - \beta } \right)   \,,  \\
  {\tau_4} =&& \left( {\frac{{3058673}}{{508032}} + \frac{{5429}}{{504}}\eta  + \frac{{617}}{{72}}{\eta ^2} - 2\sigma } \right)   \,,  \\
  {\tau_5} =&& - \left( {\frac{{7729}}{{252}} - \frac{{13}}{3}\eta } \right)\pi \,, \\
  {\tau_6} =&& - \frac{{10052469856691}}{{23471078400}} + \frac{{128}}{3}{\pi ^2} + \frac{{6848}}{{105}}{\gamma _E} \nonumber \\
             && + \frac{{3424}}{{105}}\ln {16x} + \left( {\frac{{3147553127}}{{3048192}} - \frac{{451}}{{12}}{\pi ^2}} \right)\eta \nonumber \\
             && - \frac{{15211}}{{1728}}{\eta ^2} + \frac{{25565}}{{1296}}{\eta ^3} \,, \\
  {\tau _e} =&& \frac{{785}}{{110008}} { M_c }^{{-5}/{3}} {\pi ^{{-8}/{3}}} {f_0} ^{{19}/{9}} f ^{{-43}/{9}} {e_0}^2   \,.
\end{eqnarray}
\end{subequations}
Here, $\gamma_{E} = 0.577$ is the Euler constant, $e_0$ is the eccentricity of
the binary Keplerian orbit at the fiducial frequency ${f_0}$.
Spin-orbit ($\beta $) and spin-spin ($\sigma $) parameters can be
expressed as \cite{Berti2005,PhysRevLett.74.3515} 
\begin{subequations}
\begin{eqnarray}
 \beta  =&& \frac{1}{12}\sum\limits_{i = 1}^2 {(113\, m_i^2/{M^2} + 75\, \eta )} {\bf \hat{L}} \cdot {\boldsymbol{\chi}}_i  \,, \\
 \sigma  =&& \frac{\eta }{48}( - 247\, {\boldsymbol{\chi}}_1 \cdot {\boldsymbol{\chi}}_2 + 721\, {\bf \hat L} \cdot {\boldsymbol{\chi}}_1 \, {\bf \hat L} \cdot {\boldsymbol{\chi}}_2)  \,,
\end{eqnarray}
\end{subequations}
where $\boldsymbol{\chi}_i = {\bf S}_i/m_i^2$ $(i=1,2)$ is the dimensionless spin
parameter and ${\bf S}_i$ is the spin angular momentum
for the $i$-th black hole.
${\bf \hat L}$ is the unit vector of the orbital angular momentum of the binary.
For black holes $|\boldsymbol{\chi}|  \le 1$,
$\left| \beta  \right| \lesssim  9.4$ and $\left| \sigma  \right| \lesssim  2.5$.  As an
example, if ${\bf S}_i$ aligns with ${\bf \hat L}$ and ${\chi _i} = 0.2$,
then $\beta = 1.6$ and $\sigma = 0.1$ for equal-mass binary.
$x$ is the PN parameter
\begin{equation}\label{eq:pnpara}
  x = (\pi M f)^{2/3} = (\pi M_c f)^{2/3}{\eta ^{- 2/5}}  \,.
\end{equation}

The 3PN GW strain phase evolution that includes the polarization modulation,
Doppler modulation and the
first-order eccentricity correction 
is given by
\cite{Buonanno2009,Cutler1998,Krolak1995}
\begin{eqnarray}\label{phase}
  \Psi (f) =&& 2\pi f{t_c} - {\phi _c} - \frac{\pi }{4} + \frac{3}{{128}}({M_c}\pi f)^{-5/3}
  \sum\limits_{k = 0}^6 {{\alpha _k}{x^{k/2} }} \nonumber \\
 &&-\phi_p (t(f))-\phi_D (t(f))+{\phi _e}(f)  \,,
\end{eqnarray}
with the coefficients
\begin{subequations}
\begin{eqnarray}
 {\alpha _0} =&& 1 \,, \\
 {\alpha _1} =&& 0  \,, \\
 {\alpha _2} =&& \frac{{3715}}{{756}} + \frac{{55}}{9}\eta  \,, \\
 {\alpha _3} =&& 4\beta - 16\pi  \,, \\
 {\alpha _4} =&& \frac{{15293365}}{{508032}} + \frac{{27145}}{{504}}\eta  + \frac{{3085}}{{72}}{\eta ^2} - 10\sigma \,,\\
 {\alpha _5} =&& \left(\frac{{38645}}{{756}} - \frac{{65}}{9}\eta \right) \left[ 1 + \frac{3}{2}\ln \left( \frac{x}{x_0} \right) \right]\pi \,,\\
 {\alpha _6} =&& \frac{{11583231236531}}{{4694215680}} - \frac{{640}}{3}{\pi ^2} - \frac{{6848}}{{21}}{\gamma _E} \nonumber \\
              && - \frac{{3424}}{{21}}\ln {16x} + \left( - \frac{{15737765635}}{{3048192}} + \frac{{2255}}{{12}}{\pi ^2} \right)\eta  \nonumber \\
              && + \frac{{76055}}{{1728}}{\eta ^2} - \frac{{127825}}{{1296}}{\eta ^3} \,,
\end{eqnarray}
\end{subequations}
where $x_0$ is the PN parameter (Eq.~\ref{eq:pnpara}) evaluated at the last stable orbit.

The polarization modulation is defined as
\begin{equation}\label{polarization}
 {\phi_p}(t(f)) = \arctan \frac{ - 2\cos\iota {F_ \times}(t(f))} {(1+\cos^2 \iota)F_{+}(t(f))}  \,.
\end{equation}

The motion of the TianQin detector around the heliocenter
will cause Doppler modulation on the
GW phase. Following \citep{Hu2018}, it can be expressed as
\begin{equation}
 {\phi _D}(t(f)) = 2\pi f R\sin \theta \cos \left(\frac{2\pi t(f)}{T} + \phi_0 - \phi \right)  \,.
\end{equation}
Here $\phi_0$ is the ecliptic longitude
of the satellite guiding center at $t=0$, $T$ is one sidereal year and $R=1$ AU.

Although binary orbit can be effectively circularized by the time of
coalescence in high frequencies for stellar-mass BHBs,
the orbital eccentricity may not be negligible for SMBHBs
during inspiraling in low frequencies.
Potentially, the eccentricity can be detected by space-borne detectors.
Here we adopt the first-order correction of GW phase
due to eccentricity  
\cite{Krolak1995}:
\begin{equation}
 {\phi _e}(f) =  - \frac{4239}{11696} ({M_c}\pi)^{-5/3} \frac{{f_0}^{19/9}}{f^{34/9}}{e_0}^2  \,.
\end{equation}

\section{Signal detection}\label{sec:detection}

The strain output of a detector, 
\begin{equation}
 s(t) = h(t) + n(t)
\end{equation}
consists of a time series of the GW stain signal $h(t)$ and the detector's
equivalent strain noise $n(t)$. The noise is assumed to be
sampled from a Gaussian stochastic process.
We define the noise-weighted inner product $({h_1}|{h_2})$
of two data ${h_1}(t)$ and ${h_2}(t)$ as \cite{PhysRevD.46.5236} 
\begin{equation}
 (h_1| h_2) \equiv 4 {\rm Re} \int_0^\infty {\frac{\tilde h_{1}^{*}(f){{\tilde h}_2}(f)}{S_{n}(f)}} df  \,,
\end{equation}
where ${\tilde h_1}(f)$ and ${\tilde h_2}(f)$ are the Fourier transforms of ${h_1}(t)$
and ${h_2}(t)$. ${{S_n}(f)}$ is the one-sided power spectral density (PSD)
of $n(t)$ 
for which the proposed functional form for TianQin is
provided in \cite{Luo2016,Hu2018}
\begin{equation}\label{eq:PSD}
 S_n(f)  = \frac{S_x}{L_0^2} + \frac{4S_a}{(2\pi f)^4 L_0^2}  \left( 1 + \frac{10^{-4}{\rm{Hz}}}{f} \right)   \,,
\end{equation}
where
$L_{0} = 1.73  \times 10^5 ~{\rm{km}}$ is the arm length.
$S_{x} = 10^{- 24}~{\rm{m}}^{2}/{\rm{Hz}}$ and
$S_{a} = 10^{- 30}~{\rm{m}}^{2}{\rm{s}}^{- 4}/{\rm{Hz}}$
are the PSDs of the position noise and residual acceleration noise, respectively.

The detectability of a given signal $h(t)$ is determined by the
optimal signal-to-noise ratio (SNR) \cite{PhysRevD.46.5236, 2015CQGra..32a5014M}
\begin{equation}\label{eq:SNR}
 \rho \equiv (h| h)^{1/2} =\left( \int_0^\infty \left[ \frac{h_c(f)}{h_{n}(f)} \right]^2 {\rm d}\log f \right)^{1/2}  \,,
\end{equation}
where $h_c(f)=2f|{\tilde h}(f)|$ is  the characteristic strain of the signal and
$h_n(f)=\sqrt{f S_n(f)} $ is the counterpart of the noise.
Note that both of these are dimensionless.

The average squared strain response for TianQin in Eq.~(\ref{eq:Qfac})
can be obtained by integrating over all possible sky locations
($\theta  \in [0,\pi ], \phi  \in [0,2\pi] $), polarizations
($\psi  \in [0,\pi] $), source orientations ($\iota  \in [0,\pi] $), and the time
within one orbital period of the satellites around the geocenter (${T_{\rm{sc}}} = 3.15 \times {10^5}~{\rm{s}}$):
\begin{eqnarray}
\langle |Q|^2 \rangle = && \langle {{(1 + {{\cos }^2}\iota )}^2}F_ + ^2(t,\theta,\phi,\psi ) + \nonumber \\
&&{{(2\cos \iota )}^2}F_{\times}^2(t,\theta, \phi,\psi ) \rangle  = 0.48 \,.
\end{eqnarray}
\begin{figure}
  \includegraphics[width=3.5 in]{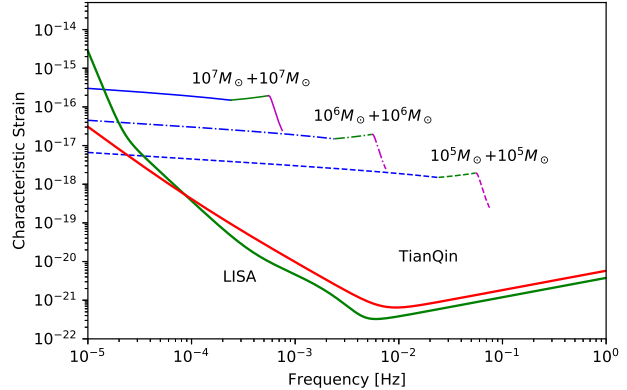}
  \caption{\label{fig:CharStrain_IMR} The characteristic strain
  of the GW signal $h_c(f)$ for three typical equal-mass
  (${10^5+10^5}{M_ \odot}$, ${10^6+10^6}{M_ \odot}$,
  ${10^7+10^7}{M_ \odot}$) SMBHB systems.
  The blue lines denote $h_c(f)$ from inspiral stage, while
  the green and purple lines from merger and ringdown stages.
  The red line and green line represent the characteristic strains of the detector
  noise $h_{n}(f)$ for TianQin and LISA \cite{2017PhRvD..95j3012B}, respectively.
  }
\end{figure}
The frequency-domain waveform (see Eq.~(\ref{eq:hf})) that adopts the average
strain response $\langle |Q|^2 \rangle^{1/2}$ is given by
\begin{equation}\label{averageh}
 \tilde h(f) =  - \sqrt {\frac{1}{{40}}} \frac {{M_c}^{5 /6}} {\pi ^{2/3}{D_L}} f^{-7/6}{e^{\mathrm{i}\Psi (f)}}  \,.
\end{equation}
Combining Eqs.~(\ref{eq:PSD}), (\ref{eq:SNR}) and (\ref{averageh}), the SNR can be
reformulated as 
\begin{equation}\label{eq:rho}
 \rho  = \frac {{M_c}^{5 /6}} {\sqrt {10} \pi ^{2/3}{D_L}} \sqrt {\int_{{f_{\rm in}}}^{{f_{\rm fin}}}
 {\frac{f^{-7/3}}{{S_n}(f)}} {\rm d} f}  \,,
\end{equation}
where ${f_{\rm fin}} = \min({f_{\rm ISCO}},{f_{\rm end}})$
with the GW frequency at the innermost stable circular orbit
${f_{\rm ISCO}} = {1/(6^{3/2}M\pi)}$ Hz
and the upper cutoff frequency for TianQin ${f_{\rm end}} = 1~{\rm{Hz}}$,
${f_{\rm in}} = \max ({f_{\rm low}},{f_{\rm obs}})$ with the
lower cutoff frequency ${f_{\rm low}} = {10^{ - 5}}~{\rm{Hz}}$
and the initial observation frequency
${f_{\rm obs}} = 4.15 \times {10^{-5}} (M_c/10^{6} ~M_{\odot})^{-5/8}(T_{\rm obs}/1~{\rm yr})^{-3/8}$ Hz.
We will choose $T_{\rm obs} = 3$ month which is the time window of
each separated observation session.
This is mainly due to the unique feature of TianQin's mission operation
which is intended to reduce the interference from the sunlight on the optical
links and simplify the thermal control of the satellites \citep{Luo2016}.

Fig.~\ref{fig:CharStrain_IMR} shows the characteristic strain of
the detector noise $h_{n}(f)$ for TianQin (red line) 
as well as the most recent one for LISA (green line) \cite{2017PhRvD..95j3012B}, and
the characteristic strains of the signal $h_{c}(f)$ for three typical equal-mass
(${10^5+10^5}{M_ \odot}$, ${10^6+10^6}{M_ \odot}$,
${10^7+10^7}{M_ \odot}$) SMBHBs located at $z=0.5$.
Here and hereafter, we refer the mass of SMBHB to the value
measured in the source rest frame considering its direct connection with
astrophysical investigations on the mass dependence of detection rates.
However, in our calculation of SNR and FIM, we take into account the
cosmological effects discussed in Sec.~\ref{subsec:twaveform}.
The blue lines with $h_{c} \propto f^{-1/6}$ present the GW contributions
from the inspiral stage of the binaries which terminates at ${f_{\rm ISCO}}$.
This is the most relevant part to the current work.
Additionally, for the purpose of illustration and comparison,
the corresponding ones from the merger (green lines)
and ringdown (purple lines) stages have also been plotted
based on the phenomenological waveform \cite{Ajith2011}.

\begin{figure}
  \includegraphics[width=3.5 in]{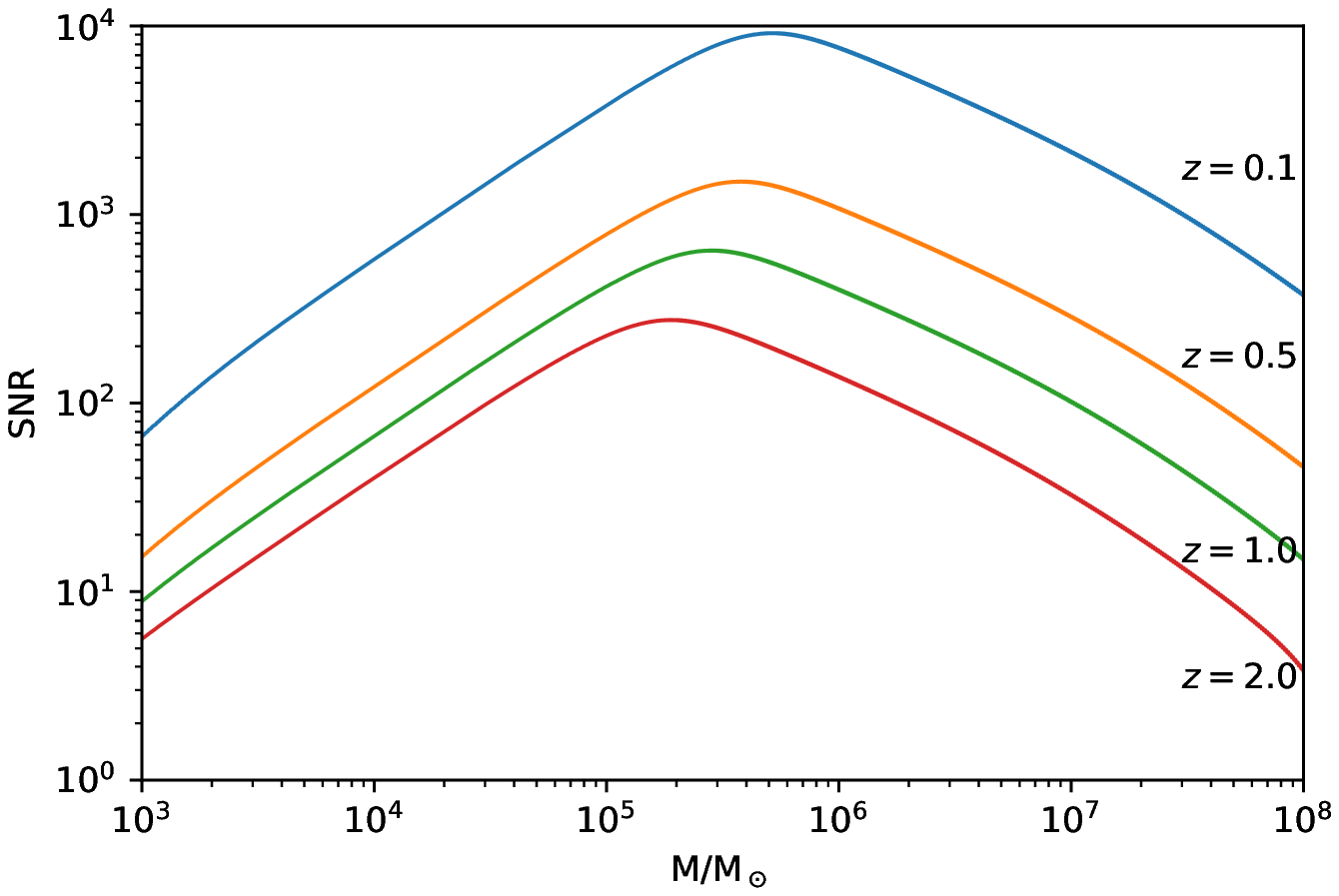}
  \includegraphics[width=3.5 in]{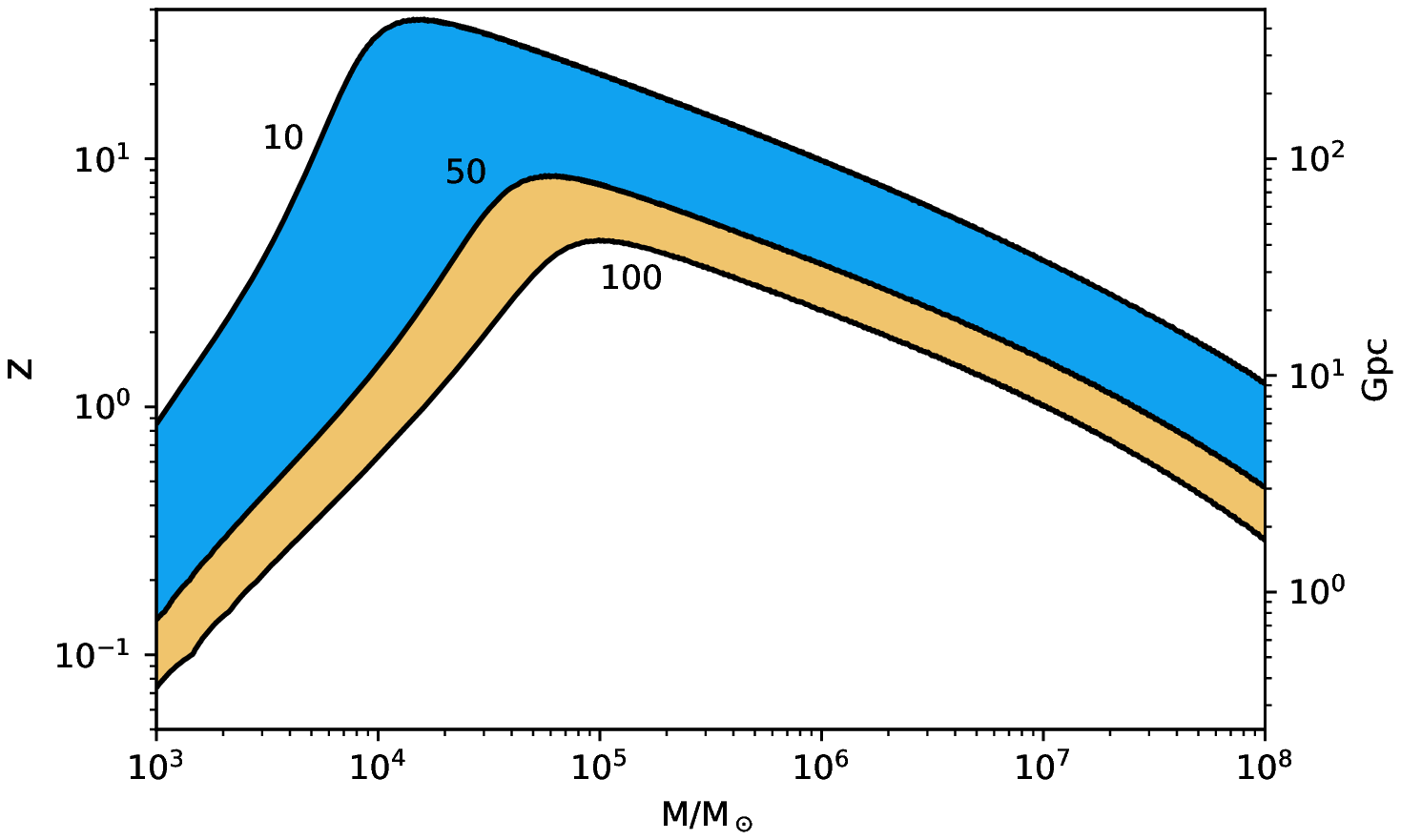}
 \caption{\label{fig:detection_horizon}
 Upper panel shows the SNR $\rho$ as a function of the total mass $M$
 for the SMBHBs located at $z=0.1$, $0.5$, $1.0$ and $2.0$ (from top to bottom).
 Lower panel shows TianQin's SMBHB horizon distance in terms of cosmological redshift
 (left vertical axis) and luminosity distance (right vertical axis) as a function
 of $M$ for fixed $\rho=10$,  $50$, and $100$.
 These results are made for the equal-mass binaries in the inspiral stage.
 The observation time $T_{\rm{obs}}$ is three months.  }
\end{figure}

For $h_{c} \propto M_{c}^{5/6} \propto M^{5/6}$, the overall amplitude of the
signal increases almost linearly with the total mass $M$,  
while the line that denotes the inspiral stage owns a lower cut-off frequency
for a more massive system (${f_{\rm ISCO}} \propto M^{-1}$).
Consequently, the competition between the integrand and
the integration limits in Eq.~(\ref{eq:rho}) will result in a maximum of $\rho$
for a given $D_L$.  For equal-mass systems, this is demonstrated in the upper panel of
Fig.~\ref{fig:detection_horizon}, in which four $D_L$ corresponding to
$z=0.1$, $0.5$, $1.0$ and $2.0$ (from top to bottom) are used.
The maxima of $\rho$ locate at $\sim$ a few $\times {10^5}{M_ \odot}$.

The lower panel of Fig.~\ref{fig:detection_horizon} shows the horizon distance
in terms of cosmological redshift (left axis) and luminosity distance (right axis) as a function of $M$ for fixed
$\rho=10$, $50$, and 100 (from top to bottom), which may be regarded respectively,
as criteria of the weak signal that can be marginally detected,
the intermediate signal, and the strong signal.
Specifically, for marginal detections with $\rho=10$, we can see that
TianQin is capable of detecting sources as far as $z > 30$ for
the lower end of the mass spectrum of
the SMBHB systems 
that are assembled in the early Universe.
The detection space shown in Fig.~\ref{fig:detection_horizon} is largely overlapped
with eLISA \cite{Hu2017,2013arXiv1305.5720E}.

\begin{figure}
  \includegraphics[width=3.5 in]{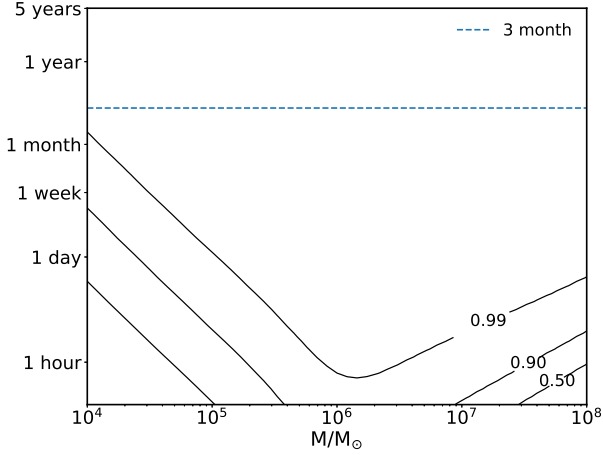}
  \caption{\label{fig:cm_t_snr_rate1}
  The contours represent $99\%$, $90\%$ and $50\%$ of the total SNR that
  can be obtained within the observation time intervals before the final merger at
  $f_{\rm ISCO}$. Equal-mass binaries and $z=1.0$ are assumed. }
\end{figure}

Fig.~\ref{fig:cm_t_snr_rate1} shows the fraction of the total SNR
that can be accumulated in the observation time intervals before
the merger of the equal-mass binary located at $z=1.0$. 
The total SNR is estimated by assuming TianQin's continuous operation of 5 years
(without the separation of $T_{\rm obs} = 3$ months observation sessions). 
The minimum of the $99\%$ of the total SNR curve is determined by the interplay between 
the time to coalescence of a system $\tau$ ($\propto M^{-5/3}f^{-8/3}\eta^{-1}$) \cite{PhysRevD.70.042001} 
and the characteristic strains of the detector noise $h_{n}(f)$ in Fig.~\ref{fig:CharStrain_IMR}. 
For $10^6+10^6 {M_ \odot}$ SMBHBs, one can see that more than 
$99\%$ of the total SNR can be obtained within one hour before the merger. 
Notice that over the vast majority of the mass range, the signal from the last few days' observation
will dominate the total SNR. Therefore, it is likely that, in practice,
the signal from a SMBHB will either be detected by TianQin
within $T_{\rm obs}$ or completely missed.


\section{Fisher information matrix}\label{sec:FIM}

Let ${\bar{\lambda}^i}$ be the true value of parameter ${\lambda^i}$
and $\hat{\lambda}^i = {\bar{\lambda}^i} + \Delta {\lambda^i}$ be the
estimated value from the data in the presence of noise.
$\hat{\lambda}^i$ can be obtained from an estimator, such as
the maximum-likelihood estimator (MLE) \cite{1993.book.....KayI}.
Suppose that the signal is sufficiently strong; the joint probability density
function for the estimation error $\Delta \boldsymbol{\lambda}$  can be
approximated by a multivariate Gaussian distribution \cite{Cutler1994,Creighton2011}
\begin{equation}
 p(\Delta \boldsymbol{\lambda}) \propto \exp ( - \frac{1}{2}{\Gamma _{ij}}\Delta {\lambda ^i}\Delta {\lambda ^j})  \,,
\end{equation}
where
\begin{equation}\label{eq:FIM}
 \Gamma _{ij} \equiv \left( \frac{\partial h}{\partial{\lambda^i}}(\hat{\lambda}^i),\frac{\partial h}{\partial {\lambda ^j}}(\hat{\lambda}^i) \right)
\end{equation}
is the $(i,j)$ entry of the \textit{Fisher information matrix} (FIM).
Particularly, the rms error of parameter ${\lambda ^i}$
\begin{equation}\label{eq:rms}
{(\Delta {\lambda^i})_{\rm rms}} = \sqrt {{\Sigma_{ii}}}  \,,
\end{equation}
where the \textit{variance-covariance matrix} (or simply the covariance matrix)
$\Sigma  = {\Gamma^{-1}}$.  
The correlation coefficient
between $\Delta \lambda^i$ and $\Delta \lambda^j$ can be defined as
\begin{equation}\label{eq:c}
 c_{ij} = \Sigma_{ij} / \sqrt{\Sigma_{ii}\Sigma_{jj}}   \,.
\end{equation}
It is a dimensionless ratio that indicates the degree to which
$\Delta \lambda^{i}$ and $\Delta \lambda^{j}$ are linearly correlated.

In this work, we focus mainly on the following parameters of the inspiraling
SMBHB signal included in Eq.~(\ref{eq:hf})  
\begin{equation}
 \boldsymbol{\lambda} = \{ \ln {M_c},\ln {D_L},\ln \eta ,{t_c},{\phi _c}, \theta, \phi, \beta ,\sigma, e_{0} \}  \,.
\end{equation}
The derivative parameters not presenting explicitly in Eq.~(\ref{eq:hf})
can be obtained by the propagation of errors. For example, the rms error
associated with the reduced mass is
\begin{equation}\label{eq:mu}
  \Delta \ln \mu  = \left[ \Sigma _{\ln {M_c}\ln {M_c}} + \frac{4}{25}{\Sigma_{\ln \eta \ln \eta }} + \frac{4}{5}\Sigma _{\ln {M_c}\ln \eta } \right]^{{1/ 2}}.
\end{equation}
In addition, the error in the sky localization $\Delta \Omega $ can be
expressed in terms of the rms errors in $\theta$ and $\phi$ \cite{Berti2005}
\begin{equation}\label{eq:angRes}
\Delta \Omega =  2\pi |\sin {\theta}| \left[ \Sigma_{\theta \theta}\Sigma_{\phi \phi} -  \Sigma^{2}_{\theta \phi}  \right]^{1/2}  \,.
\end{equation}
Note that both Eq.~(\ref{eq:mu}) and Eq.~(\ref{eq:angRes}) take into account the
correlations between parameters.

The derivatives of ${\tilde h}$ with respect to the parameters used
in Eq.~(\ref{eq:FIM}) are listed below:
\begin{subequations}\label{eq:D_Mc}
\begin{eqnarray}
 \frac{{\partial \tilde h}}{{\partial \ln {M_c}}} &&= \left( {\mathrm{i} {M_c}\frac{{\partial \Psi }}{{\partial {M_c}}}} \right)\tilde h   \,,\\
 \frac{{\partial \tilde h}}{{\partial \ln {D_L}}} &&=  - \tilde h \,,\\
 \frac{{\partial \tilde h}}{{\partial \ln \eta }} &&= \left( {\mathrm{i} \eta \frac{{\partial \Psi }}{{\partial \eta }}} \right)\tilde h \,,\\
 \frac{{\partial \tilde h}}{{\partial {t_c}}} &&= \left( {\mathrm{i}\frac{{\partial \Psi }}{{\partial {t_c}}} + \frac{1}{Q}\frac{{\partial Q}}{{\partial {t_c}}}} \right)\tilde h \,,\\
 \frac{{\partial \tilde h}}{{\partial {\phi _c}}} &&=  - \mathrm{i} \tilde h \,,\\
 \frac{{\partial \tilde h}}{{\partial {\theta}}} &&= \left( {\frac{1}{Q}\frac{{\partial Q}}{{\partial {\theta}}} + \mathrm{i} \frac{{\partial \Psi }}{\partial {\theta}}} \right)\tilde h \,,\\
 \frac{{\partial \tilde h}}{{\partial {\phi}}} &&= \left({\frac{1}{Q}\frac{{\partial Q}}{{\partial {\phi}}} + \mathrm{i} \frac{{\partial \Psi }}{\partial {\phi}}} \right)\tilde h \,,\\
 \frac{{\partial \tilde h}}{{\partial \beta }} &&= \frac{{3\mathrm{i} }}{{32}}{\eta ^{-3/5}}{\left( {\frac{{G{M_c}\pi f}}{{{c^3}}}} \right)^{-2/3}}\tilde h \,,\\
 \frac{{\partial \tilde h}}{{\partial \sigma }} &&= - \frac{{15\mathrm{i}}}{{64}}{\eta ^{-4/5}}{\left( {\frac{{G{M_c}\pi f}}{{{c^3}}}} \right)^{-1/3}}\tilde h \,,\\
 \frac{{\partial \tilde h}}{{\partial {e_0}}} &&= - \frac{{4239 \mathrm{i}}}{{5848}}{\left( {\frac{{G{M_c}\pi }}{{{c^3}}}} \right)^{-5/3}}\frac{f_0^{19/9}}{f^{34/9}}{e_0}\tilde h  \,.  
\end{eqnarray}
\end{subequations}
The explicit expressions for ${\partial \Psi}/{\partial{\lambda^i}}$
and  ${\partial Q}/{\partial \lambda^i}$ on the right-hand sides of the above
equations are too lengthy to show here.

In the following calculations, we neglect
$\partial A/\partial \ln M_c$, $\partial Q/\partial \ln M_c$, and $\partial Q/\partial \ln \eta$,
since these terms are ignorable
comparing with the contributions from the terms related to $\Psi$.
We will discuss this point with further details in Sec.~\ref{sec:conclusion}.

\section{Simulations and results}\label{sec:simulation}

In this section we use the FIM method to study the parameter estimation accuracy
of inspiraling SMBHBs for TianQin.
We assume that the strain data of the detector are from
a Michelson-type configuration which consists of two-way optical links
along two arms of the satellite constellation \cite{Hu2018}.
The functional form of instrumental noise PSD given in Eq.~(\ref{eq:PSD}) and
the associated parameter values are used for TianQin.

We carry out Monte Carlo simulations of $10^3$ SMBHBs for
mass pairs listed in the first column of Table~\ref{tab1}.
The component masses are chosen such that the average SNR
of those systems can be approximately larger than 20 (cf. Fig.~\ref{fig:detection_horizon}).
In this work, SMBHBs are located at $z=0.5$, $1.0$, and $2.0$
with the inclination angle $\cos\iota \sim\mathcal{U}(- 1,1)$ and the angular positions
of the sources $\cos{\theta} \sim\mathcal{U}(- 1,1)$ and $\phi \sim \mathcal{U}(0,2\pi)$.
Here, $\mathcal{U}(a,b)$ denotes the continuous uniform distribution between interval $[a,b]$.
We set the eccentricity of the binary ${e_0}=0.2$ at ${f_0}=10^{-4}$ Hz.
As in \cite{Cutler1994}, we are mainly concerned with the effects of the spin on the other parameters
and choose $\beta = \sigma =0$ in all cases.  

\subsection{Parameter estimation of 3PN phase including spin and eccentricity effects}\label{subsec:3PN}

Table \ref{tab1} shows the medians of the rms error distributions of the parameters
employed in the 3PN waveform including both spin and eccentricity.
Generally, the rms errors increase with the decreasing of the SNR for a specific detector. 
The peaks of the SNR 
can be read from the top panel in Fig.~\ref{fig:detection_horizon}. One can see that
the more distant sources would peak at less massive systems which merge at higher frequencies
in the source frame and are subsequently redshifted to lower frequencies in the observer's frame.

Specifically, for $z=0.5$ (corresponding to ${D_L} \approx 3~{\rm{Gpc}}$),
we can see that $\Delta \ln M_c$ 
and $\Delta \ln \eta$
can be measured to be $\sim 0.02\% - 0.7\% $ 
and $\sim 4\%  - 8\% $, respectively. 
It turns out that $\Delta \ln M_c$ is more sensitive to SNR than 
$\Delta \ln \eta$. 
$\Delta \ln D_L$ can be determined to be $ \sim 1\% - 3\% $.
Despite that $D_L$ appears in the signal model as an overall factor, its estimation error does not simply scale as ${1 /\rho }$.
This is mainly due to the correlations between $\Delta\ln {D_L}$ and $\Delta\lambda_i$
(${\Sigma _{\ln {D_L} ~\lambda_i}} \ne 0$ for $\lambda_i \neq \ln {D_L}$).
Note that it is different from the overall factor ${\cal A}$ adopted in
\cite{Berti2005,Cutler1994,Poisson1995,Arun2005} which is entirely
uncorrelated with the other parameters.

For $z=0.5$, the sky-position resolution $\Delta \Omega$, which is crucial for
multi-messenger observations, can be measured to be
$\sim 1-12~{\deg ^2}$ ($10^{-5}~{\rm{str}} \approx 1/30~{\deg ^2}$).
Furthermore, the error for the estimated time of coalescence $\Delta t_c$
is less than 20 min for all cases, which may enable TianQin
to send out prompt alerts to ground and space telescopes to search for
the potential electromagnetic counterpart within $\Delta \Omega$. 

For comparison, we calculated the SNR and parameter estimation accuracy 
for LISA based on its most recent noise curve \cite{2017PhRvD..95j3012B} (see Fig.~\ref{fig:CharStrain_IMR}). 
The medians of the SNR read 2267, 1358, and 180 for the three  
equal-mass (${10^5+10^5}{M_ \odot}$, ${10^6+10^6}{M_ \odot}$, 
${10^7+10^7}{M_ \odot}$) SMBHB systems located at $z=0.5$. 
As an example, the medians of the rms errors of $\Delta \ln M_c$, $\Delta \ln \eta$, $\Delta \ln D_L$, 
and $\Delta \Omega$ for the ${10^5+10^5}{M_ \odot}$ system 
are $\sim 0.046 \%$, $\sim 33\%$, $\sim 3.7 \%$, and $\sim 26~{\deg ^2}$, respectively. 
Note that despite higher SNR, the rms errors for LISA
are larger than the ones for TianQin. 
This is mainly due to the fact that the SMBHB accumulates the majority 
of its SNR within a few days before the merger (see Fig.~\ref{fig:cm_t_snr_rate1}); hence, it behaves 
as a relatively short-lived signal, in contrast to galactic white dwarf binaries 
or EMRIs, in the space-borne detectors. 
Therefore, the amplitude and polarization modulations (see Eq.~\ref{eq:Qfac} and Eq.~\ref{polarization}) 
of the GW 
signal, in which the information of the source is embedded \cite{PhysRevD.94.081101}, 
induced by the time varying antenna pattern 
is more significant for TianQin, since the rotation period of its constellation (3.65 day) 
is two orders of magnitude shorter than the one for LISA (one year). 
A comprehensive comparison of the two detectors, in terms of the abilities of 
detection and parameter estimation for SMBHB mergers, 
will be analyzed in a followup paper as this is beyond the scope of the current work.

\begin{table*}
\caption{\label{tab1} The medians of the SNR distributions and the rms error
distributions of parameters for 13 mass pairs,
each of which has $10^3$ trials of SMBHBs.  Sources locate at redshift
$z = 0.5$, $1.0$, and $2.0$, respectively,
with samples of inclination angle $\cos\iota \sim\mathcal{U}(- 1,1)$ and angular positions
of the sources $\cos{\theta} \sim\mathcal{U}(- 1,1)$ and $\phi \sim \mathcal{U}(0,2\pi)$. 
The 3PN waveform that includes both spin and eccentricity effects is adopted.
We set the eccentricity ${e_0}=0.2$ at ${f_0}=10^{-4}$ Hz,  $t_c = 0$ sec, $\phi_c = 0$ rad, and $\beta = \sigma =0$.
The observation time $T_{\rm obs} = 3$ months.
}
\begin{ruledtabular}
\begin{tabular}{ccccccccccccc}
($m_1,m_2$)&z&SNR&${\Delta \ln{M_c}}$&${\Delta \ln{D_L}}$ &${\Delta \ln{\eta}}$ &$\Delta {t_c}$
&$\Delta{\phi _c}$&$\Delta \beta$&$\Delta \sigma$&$\Delta {e_0}$&$\Delta \Omega$   \\
$({M_ \odot})$&&&(\%)&(\%)&(\%)&$({\rm s})$&(rad)&&&$(10^{-4})$&$(10^{-5}~{\rm str})$ \\ \hline
$(10^5,10^5)$                   &0.5 &879 &0.02 &0.96 &7.74 &5.72 &4.38 &0.22 &1.48 &2.02 &35   \\
$ $                             &1.0 &437 &0.04 &2.25 &15.4 &14.6 &8.73 &0.44 &2.96 &1.68 &205   \\
$ $                             &2.0 &198 &0.11 &5.20 &31.9 &39.1 &18.6 &0.98 &6.34 &2.85 &1062 \\
$(10^5,3 \times 10^5)$          &0.5 &917 &0.03 &0.99 &5.51 &8.22 &4.56 &0.16 &1.19 &1.49 &36   \\
$ $                             &1.0 &375 &0.06 &2.46 &11.2 &22.5 &9.33 &0.33 &2.47 &1.16 &250  \\
$ $                             &2.0 &135 &0.19 &5.50 &22.2 &73.5 &19.7 &0.84 &5.58 &1.71 &1246  \\
$(3 \times 10^5,3 \times 10^5)$ &0.5 &966 &0.04 &1.03 &5.38 &10.3 &3.32 &0.20 &1.21 &0.80 &43  \\
$ $                             &1.0 &369 &0.08 &2.50 &10.5 &27.1 &6.53 &0.42 &2.43 &0.65 &238   \\
$ $                             &2.0 &126 &0.28 &5.76 &23.2 &98.8 &15.9 &1.16 &6.28 &2.73 &1304  \\
$(3 \times 10^5,6 \times 10^5)$ &0.5 &781 &0.06 &1.09 &4.66 &15.7 &3.56 &0.22 &1.23 &0.50 &46   \\
$ $                             &1.0 &283 &0.12 &2.66 &9.27 &45.3 &7.61 &0.48 &2.66 &1.16 &275  \\
$ $                             &2.0 &102 &0.41 &6.86 &20.8 &172  &19.0 &1.38 &7.00 &5.12 &1948  \\
$(6 \times 10^5,6 \times 10^5)$ &0.5 &650 &0.08 &1.27 &4.52 &21.7 &3.42 &0.27 &1.42 &0.57 &54   \\
$ $                             &1.0 &274 &0.16 &2.68 &9.26 &58.3 &7.11 &0.57 &2.89 &1.91 &309  \\
$ $                             &2.0 &90  &0.56 &7.08 &20.6 &244  &18.6 &1.74 &8.13 &8.05 &2064  \\
$(6 \times 10^5, 10^6)$         &0.5 &583 &0.10 &1.15 &4.35 &29.9 &3.80 &0.31 &1.51 &0.99 &57   \\
$ $                             &1.0 &197 &0.22 &2.87 &8.65 &93.5 &8.15 &0.70 &3.29 &2.93 &325   \\
$ $                             &2.0 &72  &0.72 &7.46 &19.8 &365  &20.7 &1.99 &8.72 &10.9 &2374 \\
$(10^6,10^6)$                   &0.5 &529 &0.13 &1.18 &4.34 &40.1 &3.83 &0.37 &1.68 &1.50 &57  \\
$ $                             &1.0 &203 &0.26 &2.96 &8.64 &113  &7.97 &0.76 &3.48 &3.82 &361  \\
$ $                             &2.0 &73  &0.86 &7.75 &20.2 &460  &20.8 &2.29 &9.65 &13.3 &2377  \\
$(10^6,3 \times 10^6)$          &0.5 &310 &0.23 &1.63 &3.82 &123  &5.74 &0.54 &2.08 &3.27 &107  \\
$ $                             &1.0 &123 &0.45 &4.04 &7.95 &348  &11.9 &1.12 &4.25 &6.97 &712  \\
$ $                             &2.0 &39  &1.27 &11.3 &16.9 &1439 &28.9 &3.26 &10.9 &20.4 &5485  \\
$(3 \times 10^6,3 \times 10^6)$ &0.5 &274 &0.29 &1.69 &4.13 &185  &4.98 &0.70 &2.49 &4.53 &120   \\
$ $                             &1.0 &103 &0.57 &4.34 &8.08 &513  &10.1 &1.47 &5.08 &9.12 &756   \\
$ $                             &2.0 &35  &1.86 &12.4 &19.6 &2296 &27.7 &4.60 &14.5 &30.5 &7708  \\
$(3 \times 10^6,6 \times 10^6)$ &0.5 &224 &0.38 &2.08 &3.82 &336   &5.90 &0.87 &2.78 &6.08 &205   \\
$ $                             &1.0 &80  &0.73 &5.23 &7.75 &1058  &12.6 &1.90 &5.94 &12.1 &1251   \\
$ $                             &2.0 &27  &2.54 &13.9 &20.3 &4464  &33.4 &5.51 &16.4 &41.8 &9742   \\
$(6 \times 10^6,6 \times 10^6)$ &0.5 &191 &0.46 &2.22 &3.94 &493   &5.76 &1.02 &3.13 &7.45 &246  \\
$ $                             &1.0 &70  &1.04 &5.80 &8.76 &1454  &13.2 &2.46 &7.23 &17.1 &1334   \\
$ $                             &2.0 &23  &3.46 &16.2 &21.8 &7106  &32.8 &6.66 &18.5 &56.8 &12144  \\
$(6 \times 10^6, 10^7)$         &0.5 &159 &0.57 &2.48 &4.22 &827   &6.76 &1.21 &3.51 &9.31 &261   \\
$ $                             &1.0 &55  &1.45 &7.04 &9.88 &2559  &13.7 &2.52 &7.24 &23.9 &2274   \\
$ $                             &2.0 &18  &2.71 &19.0 &15.3 &11016 &38.6 &7.88 &20.8 &45.1 &18247  \\
$(10^7,10^7)$                   &0.5 &137 &0.73 &2.75 &4.81 &1161  &7.07 &1.42 &3.98 &11.9 &368    \\
$ $                             &1.0 &50  &1.46 &7.88 &8.33 &3435  &14.4 &3.01 &8.18 &24.2 &2830   \\
$ $                             &2.0 &16  &3.21 &23.6 &16.0 &15135 &44.4 &10.5 &26.8 &53.5 &28219  \\
\end{tabular}
\end{ruledtabular}
\end{table*}

For $z=1.0$ (corresponding to ${D_L} \approx 6.8~{\rm{Gpc}}$) and
$z=2.0$ (corresponding to ${D_L} \approx 16 ~{\rm{Gpc}}$),
$\Delta \ln M_c$ 
and $\Delta \ln \eta$
can be measured to be $\sim 0.04\% - 1.5\%$, 
$\sim 7.8\%  - 15\% $, and $\sim 0.1\% - 3.5\%$, 
$\sim 15\%  - 32\% $, respectively. 
The sky-position resolution $\Delta \Omega$ can be measured to be
$\sim 6.8-94~{\deg ^2}$ and $\sim 35-941~{\deg ^2}$.
The error for the estimated time of coalescence $\Delta t_c$
is $\sim 15$ sec --$1.0$ hour and $\sim 39$ sec --$4.2$ hour.
$\Delta \ln D_L$ can be determined to be $ \sim 2.3\% - 8\% $
and $ \sim 5.2\% - 24\% $. 

The estimation accuracy of the other parameters are also given in Table \ref{tab1}.
Moreover, the histograms of the rms errors for 
$\lbrace\Delta \ln M_c$, $\Delta \ln \eta$,
$\Delta \phi_c$,  $\Delta t_c$,   $\Delta \ln D_L$, and $\Delta \Omega\rbrace$
and their fitted distributions for the Monte Carlo simulation of $10^3$ trials of
${10^6+10^6}{M_\odot}$ SMBHBs located at $z=0.5$
are shown in Fig.~\ref{fig:3PN+E+S} (red lines).

\subsection{Effects of spin and eccentricity on parameter estimation}

Either spin or eccentricity correction to the phase has been
ignored for simplification in many previous works
\cite{Cutler1994,Cutler1998,Poisson1995,Berti2005,Arun2005,Arun2006}.
Here, we will show how these effects alter the parameter estimation accuracy
for the case of TianQin.
Following \cite{Lang2008},
we adopt a `special' correlation matrix in which diagonal elements are
rms errors of parameters while off-diagonal elements are correlation
coefficients (see Eq.~\ref{eq:c}) between each pair of
parameters for the 3PN phase (Table~\ref{table:3PN_corr}), 
the 3PN phase including either eccentricity (Table~\ref{table:E_corr}), 
or spin (Table~\ref{table:S_corr}).
Different from \cite{Lang2008} where the values of matrix elements
are given for a specific trial, we present the medians of the rms errors in
diagonal elements and the means of the absolute values of correlation coefficients
in off-diagonal elements for $10^3$ trials of $10^6+10^6 {M_\odot}$ SMBHBs.
In the simulations, the correlation coefficients change signs in
different trials; therefore, the absolute values allow us to take particular
note of the magnitude of correlations.  We set $z=0.5$ for all sources.

In Table~\ref{table:3PN_corr},
more than half of the correlation coefficients have values $\sim 0.2-0.5$.
Three elements have values $\sim 0.7 - 0.8$, two of which are between
$\Delta \ln D_L$ with $\Delta\theta$ and $\Delta\phi$.
The correlation between mass-related parameters $\Delta \ln M_c$
and $\Delta \ln\eta$ turns out to be strongest with coefficient value $> 0.9$.

When eccentricity of the binary orbit is considered, the correlation
between $\Delta \ln M_c$ and $\Delta \ln\eta$ is enhanced somewhat
and the rms errors of both $\Delta \ln M_c$ and $\Delta \ln\eta$ are
approximately doubled (see Table~\ref{table:E_corr}).
In contrast, the correlation between $\Delta\theta$, $\Delta\phi$, $\Delta D_L$
and the other parameters (excluding $\Delta {e_0}$) and their rms errors
remain almost the same. Hence, the 3D localization error of the source
($\Delta\theta$, $\Delta\phi$ and $\Delta \ln D_L$) are not sensitive to the
inclusion of eccentricity.

\begin{figure*}
\subfigure[]{ 
\begin{minipage}{3in}
 \includegraphics[width=3.2in]{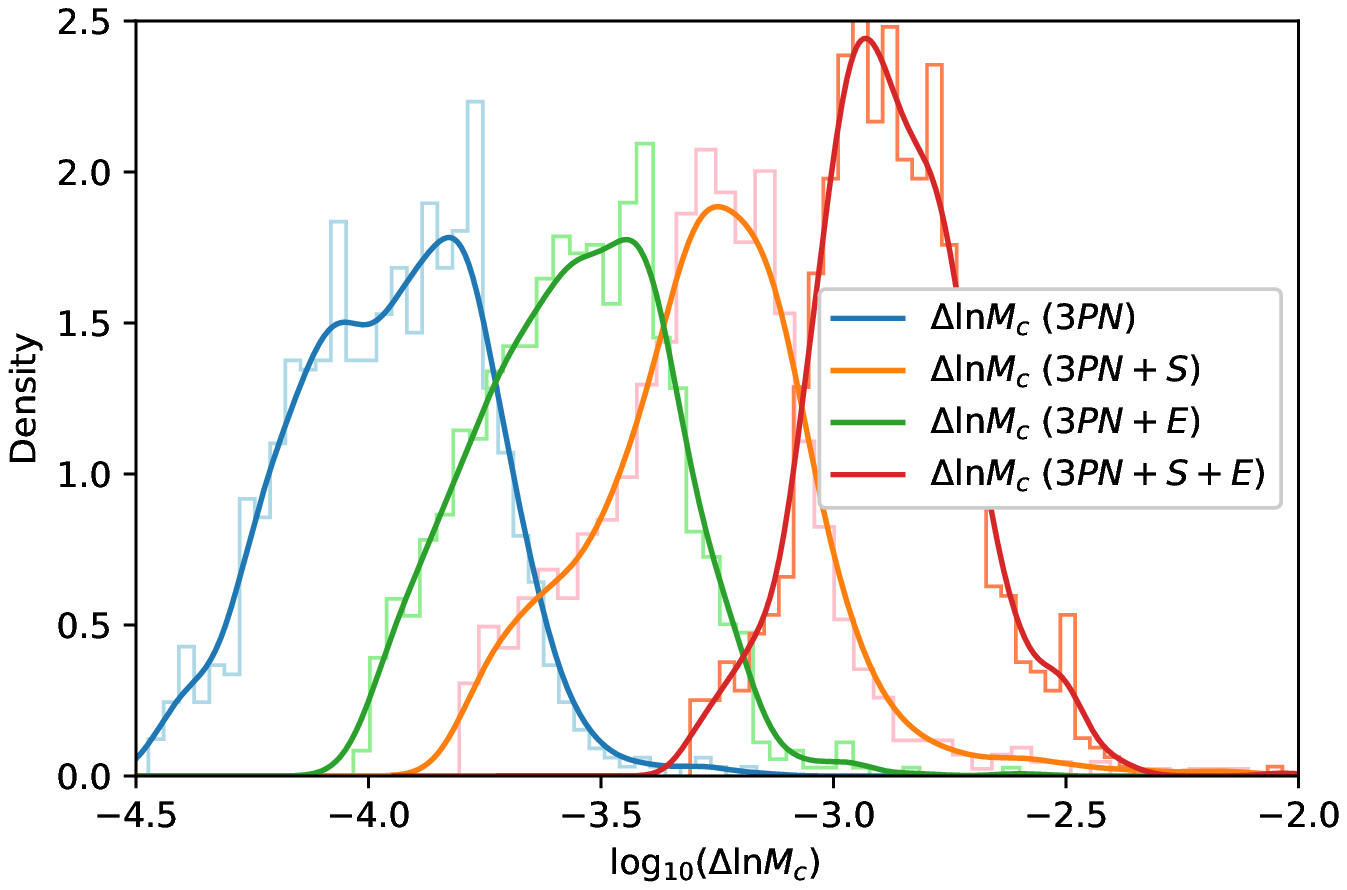}
\end{minipage}}
\subfigure[]{ 
\begin{minipage}{3in}
 \includegraphics[width=3.2in]{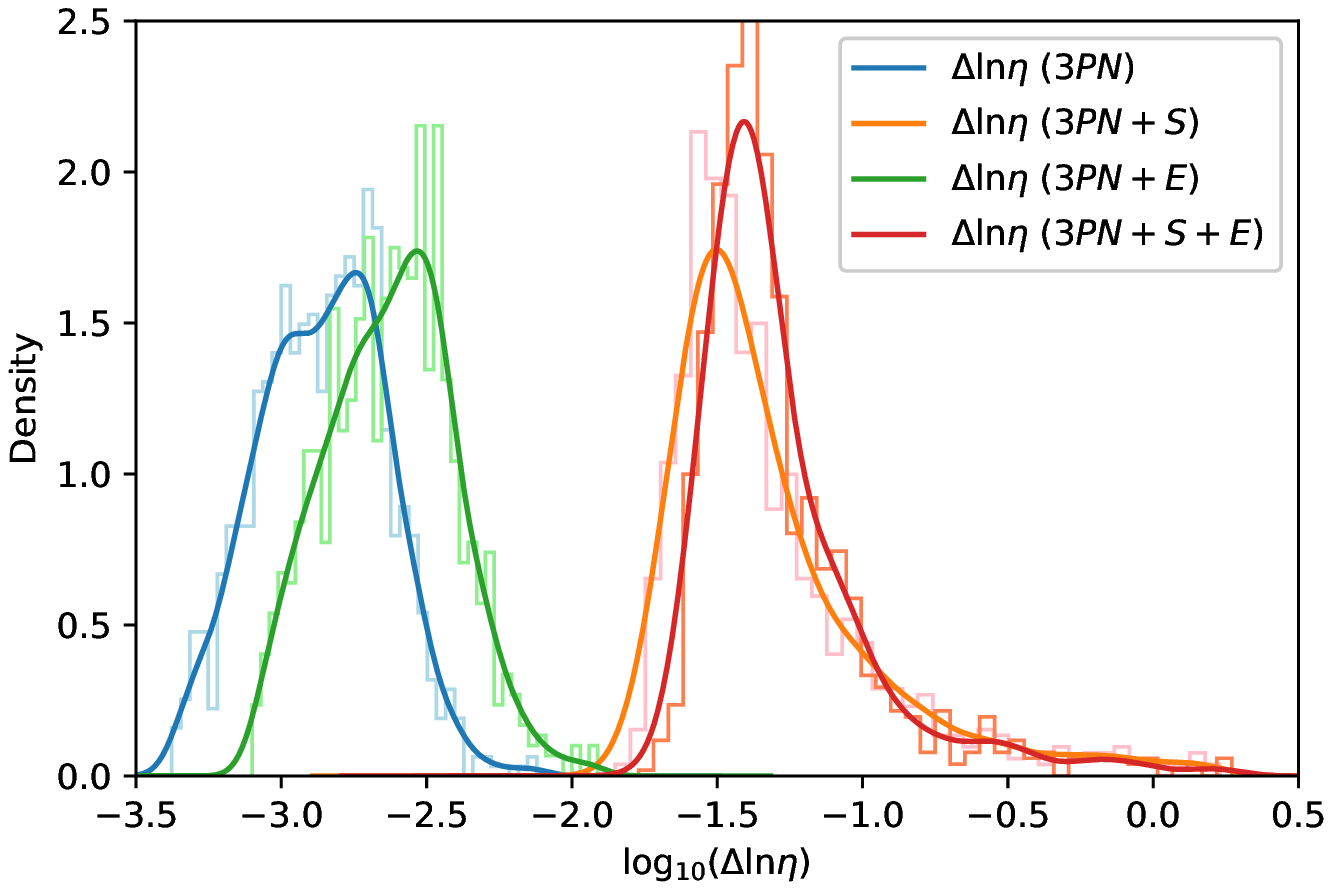}
\end{minipage}}
\subfigure[]{ 
\begin{minipage}{3in}
 \includegraphics[width=3.2in]{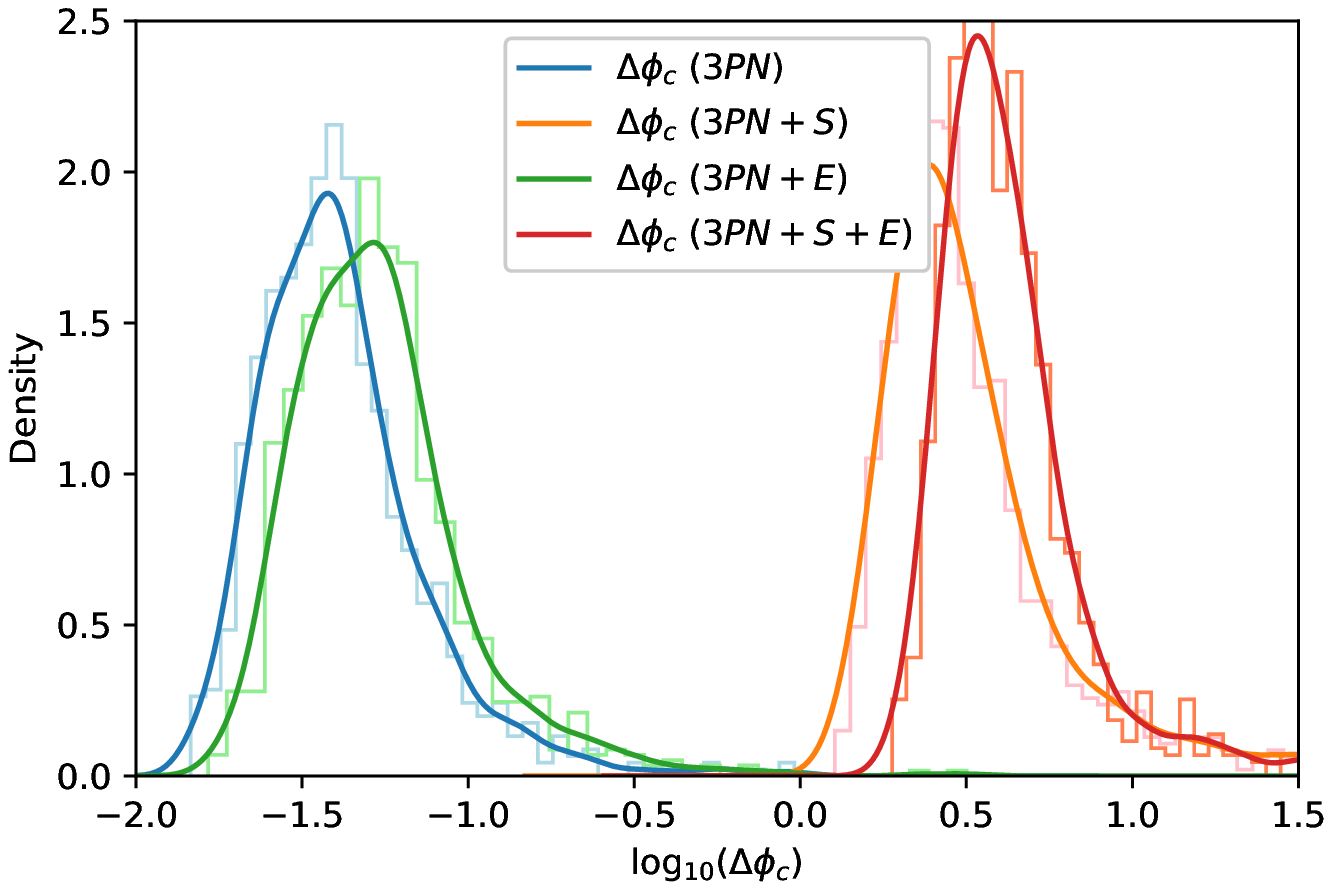}
\end{minipage}}
\subfigure[]{ 
\begin{minipage}{3in}
 \includegraphics[width=3.2in]{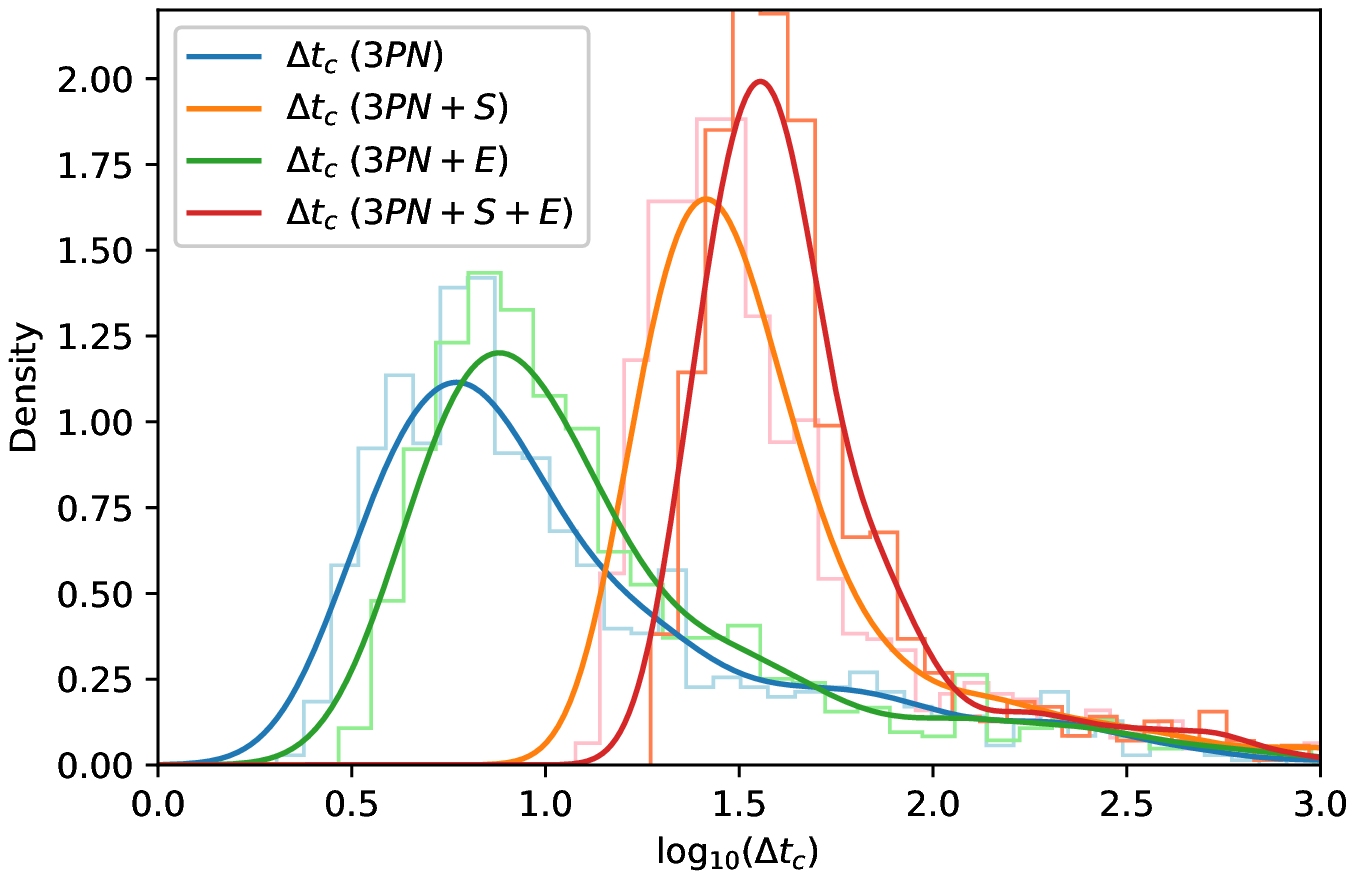}
 \end{minipage}}
 \subfigure[]{ 
 \begin{minipage}{3in}
 \includegraphics[width=3.2in]{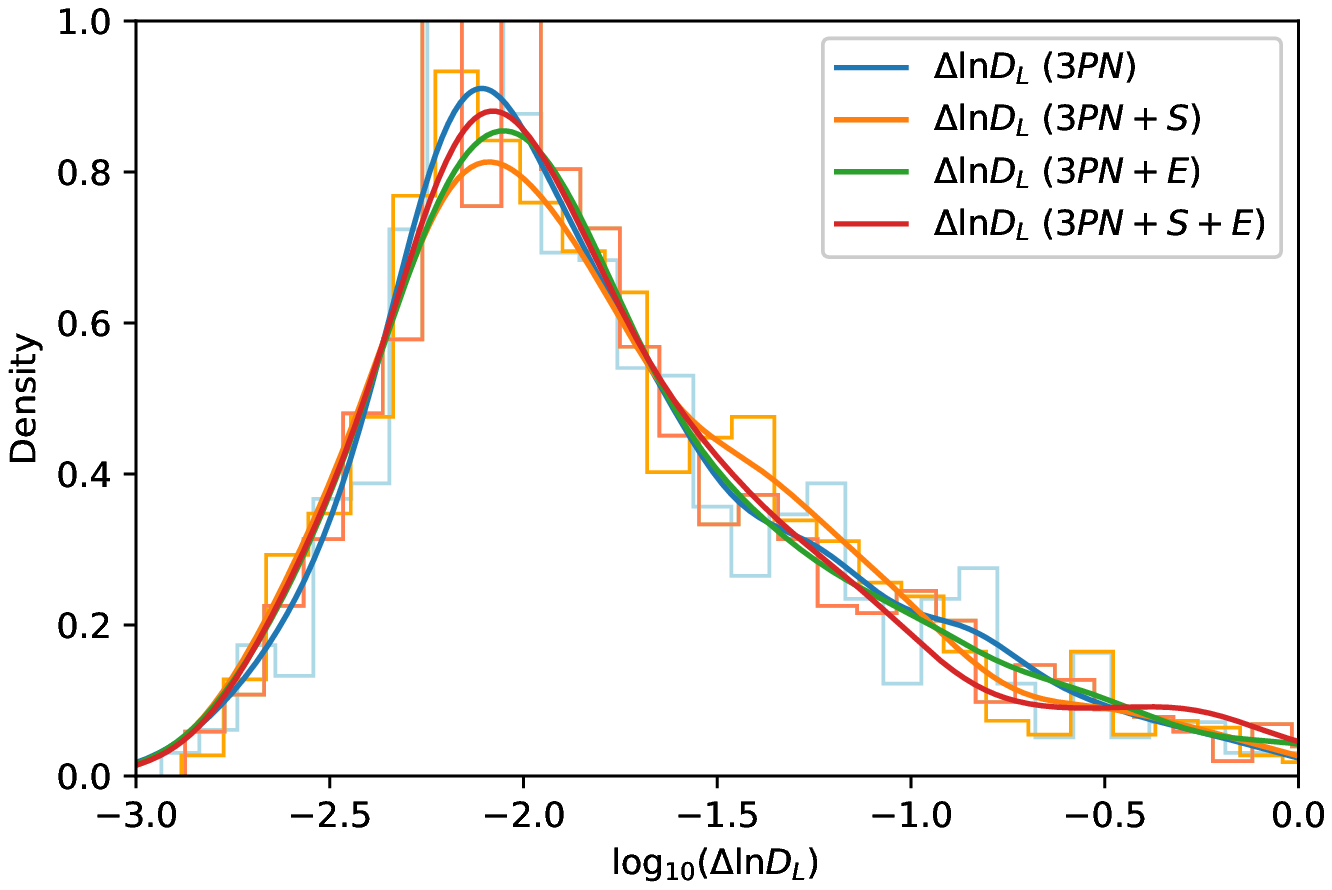}
 \end{minipage}}
 \subfigure[]{ 
 \begin{minipage}{3in}
 \includegraphics[width=3.2in]{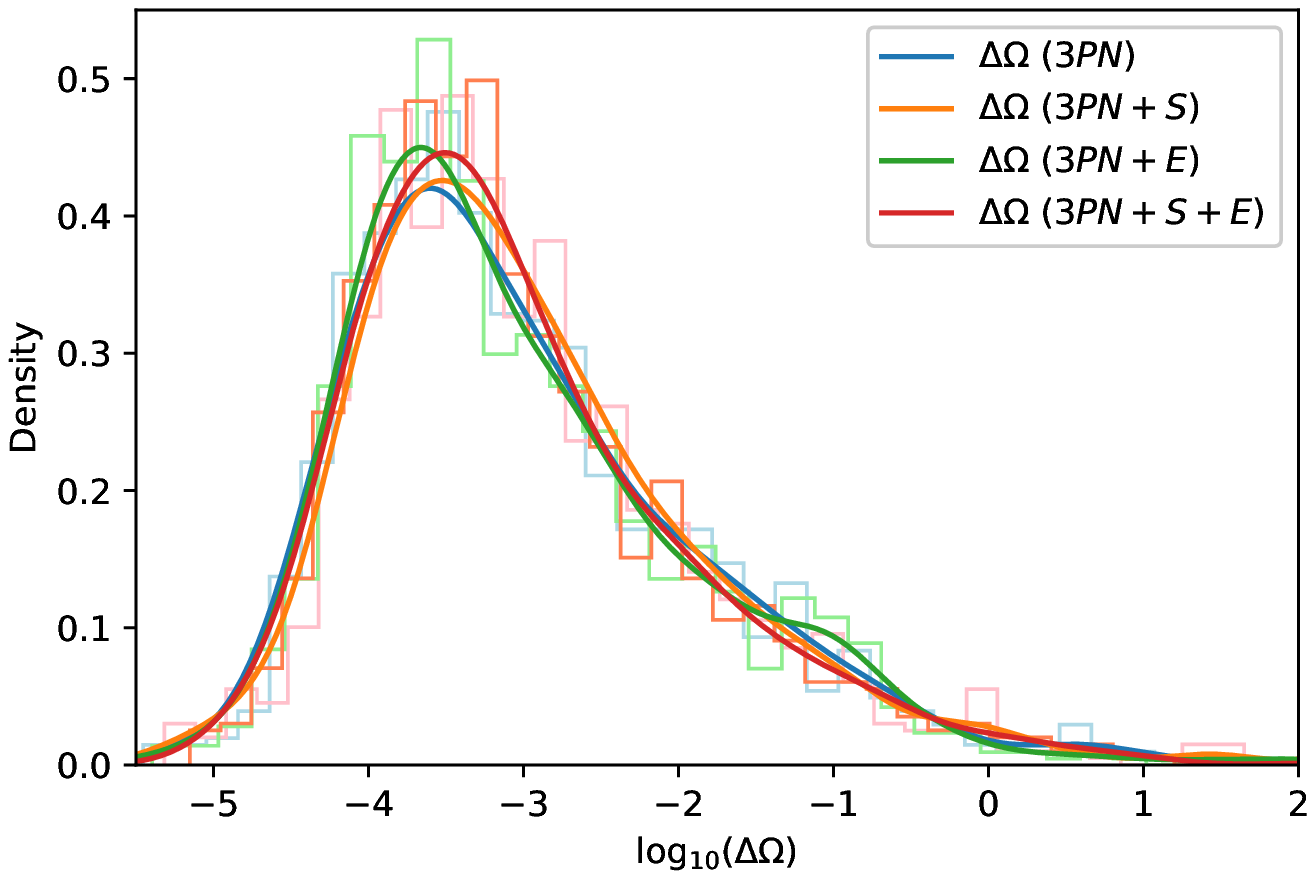}
\end{minipage}}
\caption{\label{fig:3PN+E+S} Histograms and fitted distributions of
(a) $\Delta\ln M_c$, 
(b) $\Delta \ln \eta$, (c) $\Delta \phi_c$,
(d) $\Delta t_c$, (e) $\Delta \ln D_L$ and (f) $\Delta \Omega$
for Monte Carlo simulations of $10^3$ trials of $10^6$+$10^6 {M_\odot}$
SMBHBs located at cosmological redshift $z=0.5$.
3PN, 3PN+S, 3PN+E, and 3PN+S+E denote 3PN phase (blue),
3PN phase including spin effect (orange), eccentricity effect (green)
and both (red), respectively. }
\end{figure*}

The results for 3PN phase including spin-orbit parameter $\beta$ and
spin-spin parameter $\sigma$ are given in Table \ref{table:S_corr}.
Among 36 correlation coefficients, 21 have values $\sim 0.2-0.5$,
while $\sim 1/6$ have values $0.7-0.9$.  Unlike $e_0$,
$\beta$ and $\sigma$ show strong correlation ($>0.8$) with
some of the other parameters, such as $\ln {M_c}$, $\ln \eta$,
$t_c$ and $\phi_c$, which worsens the rms error $\Delta \ln {M_c}$, $\Delta \ln \eta$,
$\Delta {t_c}$ and $\Delta {\phi _c}$ by a factor of several to
several tens ($\Delta \phi_c$ increases by a factor of $\sim 70$
considering $c_{\phi_c,\sigma}=0.986$). 
As in Table~\ref{table:E_corr}, neither the correlation coefficients relevant
to $\theta$, $\phi$ and $\ln D_L$  nor their rms errors have been changed
significantly due to the inclusion of additional parameters.
From Fig.~\ref{fig:3PN+E+S} one can see that spin (orange curves) has larger impact on
the accuracy of parameter estimation than eccentricity (green curves).
However, as noted in \cite{PhysRevD.70.042001} by using the 
\textit{simple precession} model \cite{Apostolatos1994} that the spin induced precession, 
which is neglected in the current work due to the assumption of the alignment of 
the spins to the orbital angular momentum, can significantly reduce the 
rms errors in the parameters, which is a consequence of the new signatures 
in the GW waveform introduced by precession. 
The impact of precession on parameter estimation accuracy for TianQin will be 
the subject of our future investigations.

Comparing the ${10^6+10^6}{M_\odot}$ case ($z=0.5$) in Table~\ref{tab1} with
Table~\ref{table:3PN_corr}--\ref{table:S_corr},
one can see that $\Delta \ln M_c$, $\Delta\ln\eta$, $\Delta t_c$, $\Delta\phi_c$
in the 3PN phase with eccentricity and spin (3PN+E+S)
have been worsened than the ones in 3PN, 3PN+E, and 3PN+S cases.
The correlations between mass-related parameters $(M_c,\eta)$ with $(\phi _c,t_c)$
have been discussed in \cite{Cornish2006,Lang2008,Rodriguez2013}.
As found in \cite{Lang2008}, these parameters will be underestimated if
neglecting the correlations among them.
However, $\Delta \theta$, $\Delta \phi$, $\Delta D_L$ have not been
changed significantly.
These features can be seen more evidently from Fig.~\ref{fig:3PN+E+S}.

\begin{table*}
\caption{\label{table:3PN_corr} The medians of the rms error distributions (diagonal
elements) and the mean absolute values of the correlation coefficients (off-diagonal
elements) of parameters for Monte Carlo simulation of $10^3$ trials of ${10^6+10^6}{M_\odot}$
SMBHBs  located at redshift $z=0.5$.
3PN waveform is adopted.
The rest of the setups are as in Table~\ref{tab1}.
The table is symmetric; hence, we only list the upper triangular elements.
}
\begin{ruledtabular}
\begin{tabular}{cccccccc}
&${\Delta \ln {M_c}}$&${\Delta \ln {D_L}}$&${\Delta \ln {\eta}}$&${\Delta {t_c}}$
&${\Delta {\phi _c}}$&${\Delta \theta}$&${\Delta \phi}$ \\
\hline
${\Delta \ln {M_c}}$        &0.011 &0.229 &0.907 &0.522 &0.684 &0.242 &0.273  \\
${\Delta \ln {D_L}}$        &-     &1.22  &0.262 &0.529 &0.342 &0.705 &0.704  \\
${\Delta \ln {\eta}}$       &-     &-     &0.143 &0.595 &0.807 &0.273 &0.310  \\
${\Delta {t_c}}$            &-     &-     &-     &8.06  &0.663 &0.488 &0.612  \\
${\Delta {\phi _c}}$        &-     &-     &-     &-     &0.038 &0.336 &0.395  \\
${\Delta \theta}$ &-     &-     &-     &-     &-     &0.011 &0.497  \\
${\Delta \phi}$   &-     &-     &-     &-     &-     &-     &0.015  \\
\end{tabular}
\end{ruledtabular}
\end{table*}

\begin{table*}
\caption{\label{table:E_corr}
As in Table~\ref{table:3PN_corr}, except that the 3PN waveform including first-order
eccentricity effect is adopted.
}
\begin{ruledtabular}
\begin{tabular}{ccccccccc}
&${\Delta \ln {M_c}}$&${\Delta \ln {D_L}}$&${\Delta \ln {\eta}}$&${\Delta {t_c}}$
&${\Delta {\phi _c}}$&${\Delta \theta}$&${\Delta \phi}$&${\Delta {e_0}}$ \\
\hline
${\Delta \ln {M_c}}$        &0.028 &0.255 &0.950 &0.623 &0.761 &0.265 &0.316 &0.425 \\
${\Delta \ln {D_L}}$        &-     &1.24  &0.290 &0.496 &0.323 &0.703 &0.704 &0.221 \\
${\Delta \ln {\eta}}$       &-     &-     &0.243 &0.686 &0.851 &0.303 &0.357 &0.545 \\
${\Delta {t_c}}$            &-     &-     &-     &10.1  &0.719 &0.455 &0.593 &0.435 \\
${\Delta {\phi _c}}$        &-     &-     &-     &-     &0.050 &0.330 &0.387 &0.520 \\
${\Delta \theta}$ &-     &-     &-     &-     &-     &0.010 &0.499 &0.230 \\
${\Delta \phi}$   &-     &-     &-     &-     &-     &-     &0.016 &0.271 \\
${\Delta {e_0}}$            &-     &-     &-     &-     &-     &-     &-     &0.195 \\
\end{tabular}
\end{ruledtabular}
\end{table*}

\begin{table*}
\caption{\label{table:S_corr}
As in Table~\ref{table:3PN_corr}, except that the 3PN waveform including
spin-orbit ($\beta$) and spin-spin ($\sigma$) effects is adopted.
}
\begin{ruledtabular}
\begin{tabular}{cccccccccc}
&${\Delta \ln {M_c}}$&${\Delta \ln {D_L}}$&${\Delta \ln {\eta}}$&${\Delta {t_c}}$
&${\Delta {\phi _c}}$&${\Delta \theta}$&${\Delta \phi}$
&${\Delta {\beta}}$&${\Delta {\sigma}}$ \\
\hline
${\Delta \ln {M_c}}$        &0.054 &0.198 &0.330 &0.647 &0.581 &0.205 &0.254  &0.821 &0.671 \\
${\Delta \ln {D_L}}$        &-     &1.26  &0.192 &0.343 &0.208 &0.702 &0.692  &0.203 &0.212 \\
${\Delta \ln {\eta}}$       &-     &-     &3.770 &0.589 &0.904 &0.267 &0.196  &0.433 &0.831 \\
${\Delta {t_c}}$            &-     &-     &-     &31.0  &0.791 &0.364 &0.407  &0.590 &0.830 \\
${\Delta {\phi _c}}$        &-     &-     &-     &-     &2.812 &0.270 &0.242  &0.543 &0.986 \\
${\Delta \theta}$ &-     &-     &-     &-     &-     &0.012 &0.506  &0.228 &0.264 \\
${\Delta \phi}$   &-     &-     &-     &-     &-     &-     &0.018  &0.256 &0.256 \\
${\Delta {\beta}}$          &-     &-     &-     &-     &-     &-     &-      &0.250 &0.605 \\
${\Delta {\sigma}}$         &-     &-     &-     &-     &-     &-     &-      &-     &1.131 \\
\end{tabular}
\end{ruledtabular}
\end{table*}

\subsection{Effects of PN order on parameter estimation}

The 2PN phase has been adopted as a benchmark in many previous investigations on parameter
estimation accuracy of coalescing binaries for either LIGO-Virgo type detectors \cite{Poisson1995,Krolak1995}
or LISA \cite{Berti2005,Lang2006}.
In this part, we will show the difference of the  parameter estimation accuracy forecasted
by the 2PN and 3PN phases in the context of TianQin.
For a comprehensive study, we include eccentricity and spin parameters
in the waveforms. The remaining setups of the simulations for the 2PN phase are
the same as in Sec.~\ref{subsec:3PN}, except $z=0.5$ for all sources in this subsection.

\begin{figure*}
\includegraphics[width=7.6in]{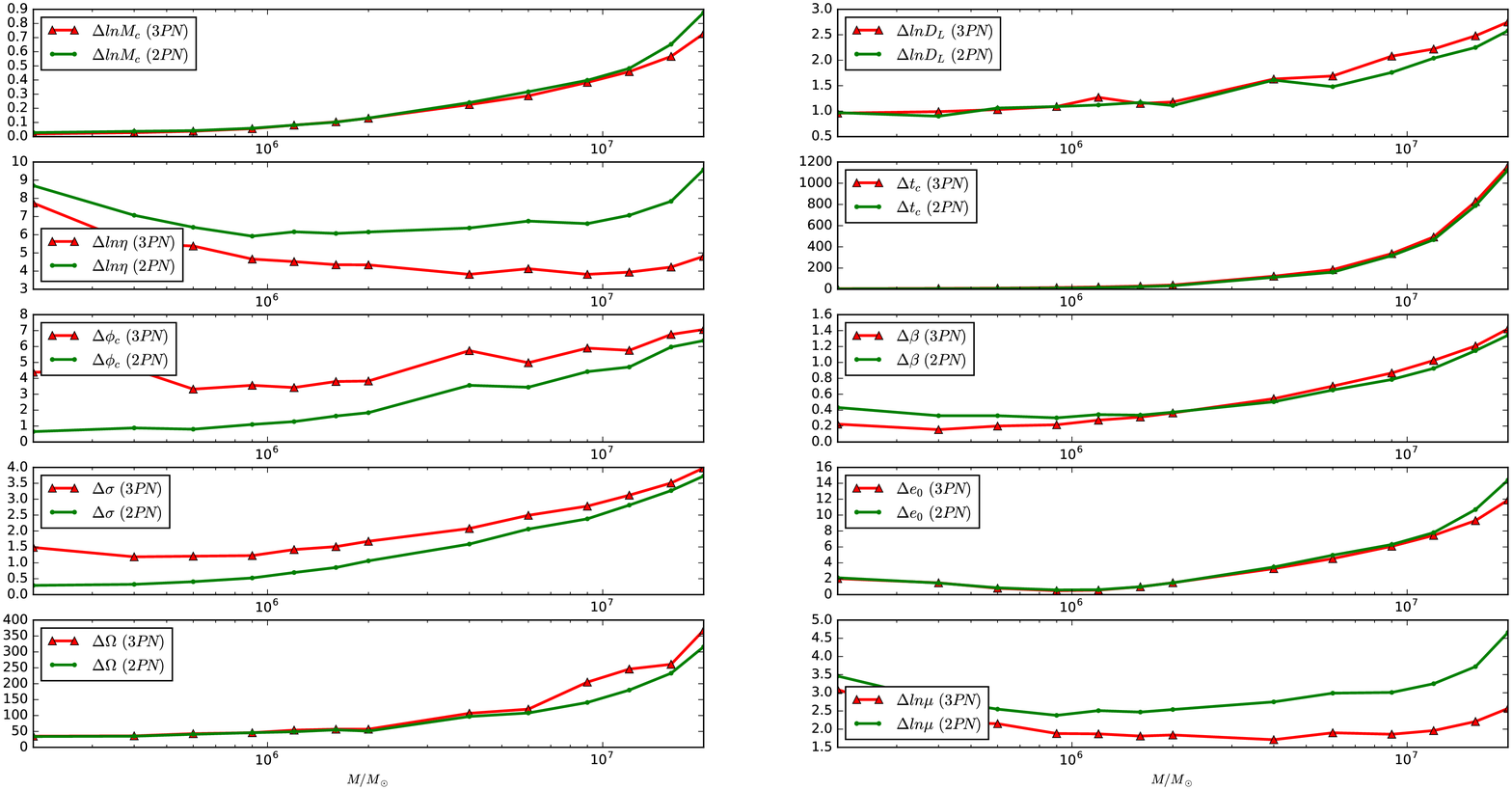}
\caption{\label{fig:ErrorsTable_3PN+2PN} Parameter estimation accuracy
 as a function of the total mass of SMBHBs. 
The 2PN and 3PN phases with both spin and eccentricity effects are considered. 
 The units of the parameters are identical to the ones used in Table~\ref{tab1}. }
\end{figure*}

Figure~\ref{fig:ErrorsTable_3PN+2PN} shows the dependencies of the medians of
the rms errors on the total mass of SMBHBs when considering the 2PN (green lines with dots)
or 3PN (red lines with triangles) phases, respectively. The components are taken from
the mass pairs listed in Table~\ref{tab1}.
We can see that the medians of $\Delta\ln M_c$, $\Delta\ln D_L$,
$\Delta e_0$, and $\Delta \Omega$ based on the 2PN phase overlap with the 3PN phase
very well in most of the region that covers the lower mass end; whereas discrepancies
between the 2PN and 3PN results, within a fraction of the respective values,
are shown in these parameters at the higher mass end ($\sim {10^7}{M_\odot}$). 
The 2PN phase overestimates the medians of $\Delta\ln\eta$ and $\Delta\ln\mu$ by a factor of 2
around the higher mass end, while it underestimates the medians of $\Delta \phi_c$
and $\Delta\sigma$ by a factor of a few around the lower mass end.
The 2PN can either underestimate or overestimate $\Delta\beta$ depending
on the value of total mass by, at most, a factor of 2. It turns out that
$\Delta t_c$ is not sensitive to the PN order at all in the concerned total mass range.

\section{Conclusions}\label{sec:conclusion}

This work studied the performance on detection and
parameter estimation accuracy of SMBHB inspirals for TianQin,
a space-borne GW detector working in the millihertz frequency band.
By calculating average SNR, we found that TianQin is capable of detecting
SMBHBs assembled in the early Universe.
The maximum distance reach in terms of cosmological redshift $z > 30$.
Using the FIM method, we calculated the estimation accuracies of the parameters
contained in the GW strain signal model.
We adopted the `restricted' 3PN frequency-domain waveform
and considered the contributions from spin (spin-orbit and spin-spin) and
first-order eccentricity effects to the phase of the GW signal.
Using Monte Carlo simulations of $10^3$ binaries sampled uniformly
in sky location and orientation,
we calculated the rms error distributions of parameters and their correlation
coefficients for SMBHBs with component masses in the range of $(10^{5},10^{7}){M_ \odot }$.
Furthermore, we studied the effects of eccentricity, spin and PN order on the parameter
estimation accuracy for TianQin.

Two sets of codes in \textsf{Python} \cite{python} and
\textsf{Mathematica} \cite{mathematica} have been developed independently
to implement the FIM method. The results from these are consistent
with each other within $1\%$ in terms of relative differences of the rms errors.
For the sake of computation time, we ignored the terms, when
evaluating the elements in Eq.~(\ref{eq:D_Mc}),
of the derivatives of the amplitude with respect to the coalescence time $t_c$,
chirp mass $M_c$ and symmetric mass ratio $\eta$ considering that they are
negligible compared to the derivatives of the phase.
To validate this approximation, we used the complete terms
including the derivatives of both amplitude and phase to calculate FIM
for a subset of the simulations in \textsf{Mathematica} codes,
from which the parameter estimation accuracies are consistent with the ones
obtained from \textsf{Mathematica} and \textsf{Python} codes that exclude
the minor terms.

Simplified assumptions used in this work can be improved in a more
careful treatment. Such as, we only took into account the first-order
eccentricity contribution to the phase but not to the amplitude.
Especially, the latter will become more important when the eccentricity
is noticeably larger and enhance significantly the amplitude around the
time of periastron. 
Besides, higher-order eccentricity
contribution \cite{Moreno-Garrido1995,Mikoczi2010,Hinder2017} to both phase and amplitude
may be introduced in the signal model.
Moreover, we assumed that the spin parameters are constant during
the inspiraling stage of SMBHBs.
It may not be true in the real situation when the spin, by chance, 
is not parallel to the orbital angular momentum; thus, procession
may be induced in orbital dynamics \cite{Apostolatos1994,PhysRevD.70.042001}.
These factors may change the results presented here considerably.

Finally, all the results presented in this work assume
a single Michelson-type interferometer configuration.
In fact, the space-borne
GW detector formed by a nearly equilateral triangular constellation  with two-way optic
links along each arm is equivalent to two \textit{independent} Michelson-type interferometers at the 
frequencies lower than the \textit{transfer frequency} ($f_{\ast}\approx 0.28$ Hz for TianQin) \cite{Vallis2009}.
For two (or more) interferometers with independent noise, the total SNR is the root of
quadratic sum of individual SNRs while the total FIM
is the sum of the individual FIMs.
We expect that the results from two interferometers will be improved
somewhat over the current ones.
However, a thorough investigation of this aspect, especially if
involving much more sophisticated data combinations from time delay
interferometry (TDI) \cite{2005LRR.....8....4T} (which is used to subtract the dominating
laser phase noise), is out of the scope of the current paper.
All the considerations mentioned above will be subject to our
further investigations for TianQin.

\begin{acknowledgments}
Y.W. is supported by the National Natural Science Foundation of
China under Grants No. 91636111, No. 11690021, and No. 11503007. 
Y.M.H. acknowledges the support from the National Natural Science
Foundation of China under Grant No. 11703098. 
This work is partly supported by ``the Fundamental Research Funds for the 
Central Universities" under Grant No. 2019kfyRCPY106. 
The authors thank the anonymous referee for helpful comments and suggestions.

\end{acknowledgments}

\appendix

\section{Frequency-domain waveform in SPA}\label{sec:SPA}

The SPA gives the leading asymptotic behavior of the generalized Fourier
integrals in the following form \cite{Bender1978}
\begin{equation}
 I(\lambda ) = \int_a^b {f(t)} {e^{\mathrm{i}\lambda g(t)}}dt  \,,
\end{equation}
where $f(t)$, $g(t)$, $a$, $b$, and $\lambda$ are all real.
A point $c \in (a,b)$ is called a \textit{stationary} point of $g(t)$
if $g'(c) = 0$. Suppose $f(c) \ne 0$, $g'(t) \ne 0$ everywhere else for
$t \in (a,b)$ and $g(t)$ is smooth enough to be expanded as a Taylor series.
The leading contribution to $I(\lambda)$ comes from a small
interval of width $\varepsilon $ surrounding the stationary
point  $c$ of $g(t)$, such that
\begin{equation}
 I(\lambda ) \approx \int_{c - \varepsilon }^{c + \varepsilon } {f(t){e^{\mathrm{i}\lambda g(t)}}dt}  \,,
\end{equation}
for $\lambda  \to  + \infty$. To obtain the leading behavior of the integral, we replace
$f(t)$ by $f(c)$ and $g(t)$ by $g(c) + {{{g^{(p)}}(c){{(t - c)}^p}} /{p!}}$, 
where ${{g^{(p)}}(c)} \ne 0$ but $g'(c) =  \cdots  = {g^{(p - 1)}}(c) = 0$.
Further, we let $s=t-c$ and replace $\varepsilon $ by $\infty$ approximately, then
\begin{equation}
 I(\lambda ) \approx 2f(c){e^{\mathrm{i}\lambda g(c)}}\int_0^\infty  {\exp [\mathrm{i}\lambda {{{g^{(p)}}(c){s^p}} /{p!}}]ds}  \,.
\end{equation}
To evaluate the integral, we rotate the contour of integration
from the real $-s$ axis by an angle $\pi/2p$ if ${g^{(p)}}(c) > 0$
and make the substitution
\begin{equation}
 s = {e^{\mathrm{i}{\pi / {2p}}} }{\left[ {\frac{{p!u}}{{\lambda {g^{(p)}}(c)}}} \right]^{1/p}}  \\,
\end{equation}
where $u$ is real. Or rotate the contour by an angle
$-\pi / 2p$  if $g^{(p)}(c) < 0$ and make the substitution
\begin{equation}
 s = {e^{-\mathrm{i} {\pi / {2p}}}}{\left[ {\frac{{p!u}}{{\lambda |{g^{(p)}}(c)|}}} \right]^{1/p}}.
\end{equation}
Thus, 
\begin{equation}
 I(\lambda ) \approx 2f(c){e^{\mathrm{i}\lambda g(c) \pm {{\mathrm{i}\pi } / {2p}}}}{\left[ {\frac{{p!}}{{\lambda |{g^{(p)}}(c)|}}} \right]^{1/p}}\frac{{\Gamma ({1 / p})}}{p}
\end{equation}
for $\lambda \to + \infty$. Here the $\Gamma(\cdot)$ is Gamma function.
For our case, $p=2$ and $\Gamma (1/2) = \sqrt \pi$, so to the leading order
\begin{equation}\label{eq:spahf}
 I(\lambda ) \approx f(c){e^{\mathrm{i}\lambda g(c) \pm {{\mathrm{i}\pi }/4}}}{\left[ {\frac{{2\pi }}{{\lambda |g''(c)|}}} \right]^{1/2}},
\end{equation}
where we use the term $+\pi/4$ if $g''(c) > 0$, or $-\pi/ 4$  if $g''(c) < 0$.

Given a GW signal $h(t) = A(t)\cos \Phi (t)$, its Fourier transform is
\begin{equation}
 \tilde h(f) = \int_{ -\infty }^\infty  {{e^{\mathrm{i} 2\pi ft}}} h(t)dt = {I_1}( f ) + {I_2}( f )
\end{equation}
where
\begin{subequations}
\begin{eqnarray}
 {I_1}( f ) &=& \frac{1}{2}\int\limits_{ - \infty }^\infty  {A( t )} {e^{\mathrm{i}[{2\pi ft - \Phi ( t )} ]}}dt   \,, \\
 {I_2}( f ) &=& \frac{1}{2}\int\limits_{ - \infty }^\infty  {A( t )} {e^{\mathrm{i}[{2\pi ft + \Phi ( t )} ]}}dt  \,.
\end{eqnarray}
\end{subequations}
Using integration by parts for $I_2$, we obtain
\begin{eqnarray}
 && 2{I_2}(f) = \frac{A(t){e^{\mathrm{i}[2\pi ft + \Phi (t)]}}}{\mathrm{i}(2\pi f +{d\Phi } /{dt} )}|_{ -\infty }^\infty   \nonumber  \\
 &&+ \int_{ - \infty }^\infty \mathrm{i}{e^{\mathrm{i}[2\pi ft + \Phi (t)]}}\frac{d}{{dt}} {\left[ \frac{A(t)} {(2\pi f + {d\Phi } /{dt} )} \right]} dt  \,,
\end{eqnarray}
where the first term on the right hand side vanishes since $h(t)=0$
when $t \to \pm \infty$.  For the second term, we introduce
Riemann-Lebesgue lemma: $\int_a^{b} f(t)e^{\mathrm{i}\lambda t} dt \to 0$
$(\lambda  \to  +\infty)$ provided that $\int_a^b {\left| f(t) \right| dt}$
exists \footnote{This condition is always satisfied since the signal
we considered is nonzero only for a finite period of time.},
such that the second term will also vanish if the following
inequality is satisfied
\begin{eqnarray}\label{eq:appcon}
  \left| \frac{d}{dt}\left[ \frac{A(t)}{2\pi f + {d\Phi }/{dt} } \right] \right|
   =&& \left| \frac{ A(t) \left[ \frac{d\ln A}{dt} (2\pi f + \frac{d\Phi }{dt}) -
   \frac{{d^2}\Phi}{d{t^2}} \right]} {( 2\pi f + \frac{d\Phi }{dt} )^2} \right| \nonumber\\
  &&\ll {M_b}   \,,
\end{eqnarray}
where $M_b$ is the maximal finite boundary value. Note that the constraints
$d\ln A / dt \ll d\Phi/dt $ and ${d^2}\Phi / {d{t^2}} \ll (d \Phi /dt)^2$ \cite{Creighton2011}
are sufficient but not necessary to make the second term vanished.
In Appendix~\ref{sec:test},
we will demonstrate that Eq.~\ref{eq:appcon} is satisfied for TianQin.
Thus the GW signal in frequency domain is
\begin{equation}
 \tilde h(f) \approx \frac{1}{2}\int_{-\infty}^{\infty} {A(t){e^{\mathrm{i}[2\pi ft - \Phi(t)]}}dt}  \,.
\end{equation}

Let $\lambda  = {10^8} \times 2\pi f$ in order to use SPA, so the
condition $\lambda  \to + \infty$ (actually sufficiently large) can always hold in the millihertz frequency
band of a space-borne GW detector, such as LISA and TianQin.
Suppose $g(t) = {10^{ - 8}}[t - {\Phi (t)} /(2\pi f)]$,
then $g'(t) = {10^{ - 8}}[1 - {\Phi '(t)} /(2\pi f)]$,
thus the stationary point $t_{\rm{sp}}$ of $g(t)$ is the time at which $d{\Phi (t)}/ {dt} = 2\pi f$.
Furthermore, for the second derivative $g''(t) =  - {10^{ - 8}}{\Phi ''(t)} / (2\pi f) < 0 $,
we use the factor $e^{- i\pi /4}$.

Using Eq.~\ref{eq:spahf}, we can obtain $\tilde h(f)$ for $f>0$:
\begin{eqnarray}
   \tilde h(f) &&\approx \frac{1}{2}A({t_{\rm{sp}}}){e^{\mathrm{i}[\lambda g({t_{\rm{sp}}}) - \pi /4]}}\sqrt {\frac{{2\pi }}{{\lambda |g''({t_{\rm{sp}}})|}}}  \nonumber\\
    &&= \frac{1}{2}A({t_{\rm{sp}}}){\left(\frac{df}{dt}\right)_{\rm{sp}}^{-\frac{1}{2}}}{e^{\mathrm{i}(2\pi f{t_{\rm{sp}}} - \Phi ({t_{\rm{sp}}}(f)) - \pi /4)}}  \,.
\end{eqnarray}

\section{Validation of SPA for TianQin}\label{sec:test}

The validity of SPA depends on the satisfaction of Eq.~(\ref{eq:appcon}) 
in which the three important terms are the derivatives of amplitude
and phase in the time domain with respect to $t$. 
We set $t_c = \phi_c = \psi  = \iota  = 0$, $\beta=\sigma=0$,
and $e_0=0$ in the waveforms.
The expressions of the amplitude and the phase of the waveform in
the time domain are given by
\begin{widetext}
\begin{eqnarray}
A(t,\theta , \phi ) &=&  - \frac{{{M_c}}}{{2{D_L}}}{\left( {\frac{{{t_c} - t}}{{5{M_c}}}} \right)^{ - 1/4}}Q(t, \theta ,\phi )   \,, \\
Q(t,\theta , \phi ) &=& \sqrt {({{(1 + \cos {\iota ^2})}^2}{F_ + }{{(t, \theta , \phi )}^2} + {{(2\cos \iota )}^2}{F_ \times }{{(t, \theta , \phi )}^2})}   \,,  \\
\Phi (t,\theta , \phi ) &=& {\phi _c} - \frac{{2{\Theta ^{5/8}}}}{\eta }\left( {1 +\sum\limits_{k = 1}^5 {{a_i}{\Theta ^{ - (i+1)/8}}}} \right)+{\phi_p}(t)+{\phi_D}(t)   \,,
\end{eqnarray}
\end{widetext}
with \cite{Creighton2011}


\begin{subequations}
\begin{eqnarray}
 \Theta  &=& \frac{{\eta ({t_c} - t)}}{{5M}}  \,,  \\
 {a_1} &=& \left( {\frac{{3715}}{{8064}} + \frac{{55}}{{96}}\eta } \right)   \,,  \\
 {a_2} &=& \frac {-3 \pi}{4}\,, \\
 {a_3} &=& \left( {\frac{{9275495}}{{14450688}} + \frac{{284875}}{{258048}}\eta  + \frac{{1855}}{{2048}}{\eta ^2}} \right)   \,.\\
 {a_4} &=& \left(\frac{{-38645}}{{172032}} + \frac{{65}}{2048}\eta \right) \ln \left( \frac{\Theta}{\Theta _0} \right)\pi \,,\\
 {a_5} &=& \frac{{831032450749357}}{{57682522275840}} - \frac{{53}}{40}{\pi ^2} - \frac{{107}}{{56}}{\gamma _E} \nonumber \\
              && + \frac{{107}}{{448}}\ln \left( \frac{\Theta}{256} \right) + \left( - \frac{{126510089885}}{{4161798144}} + \frac{{2255}}{{2048}}{\pi ^2} \right)\eta  \nonumber \\
              && + \frac{{154565}}{{1835008}}{\eta ^2} - \frac{{1179625}}{{1769472}}{\eta ^3}\,.
\end{eqnarray}
\end{subequations}

Taking the typical ${10^6}+{10^6} {M_ \odot }$ binary system with cosmological
redshift $z=0.5$ as an example,  we evaluated the left-hand side of Eq.~(\ref{eq:appcon}) for TianQin
and found that its value is well bounded for systems located in $\theta  \in (0,\pi )$ and $\phi  \in (0,2\pi )$
at the concerned time interval $t \in ({t_{\rm{in}}},{t_{\rm{fin}}})$. Here ${t_{\rm{in}}}$ and ${t_{\rm{fin}}}$
are the time that $f={f_{\rm in}}$ and  ${f_{\rm fin}}$.


%

\end{document}